\documentclass[aps,english,twocolumn,preprintnumbers,superscriptaddress,showkeys,showpacs]{revtex4}
\usepackage{subfigure}
\usepackage{graphicx}
\usepackage{amssymb}
\usepackage{amsmath}
\usepackage{color}

\makeatletter

\usepackage{babel}
\makeatother

\begin{document}
\preprint{HIZ15, dated: \today}
\title{Chimera states in population dynamics: networks with fragmented and hierarchical connectivities}

\author{Johanne Hizanidis}
\email[corresponding author: ]{hizanidis@physics.uoc.gr}
\affiliation{Institute of Nanoscience and Nanotechnology, National Center for Scientific Research ``Demokritos'', 15310 Athens, Greece}
\affiliation{Crete Center for Quantum Complexity and Nanotechnology, Department of Physics,
University of Crete, 71003 Heraklion, Greece}

\author{Evangelia Panagakou}
\affiliation{Institute of Nanoscience and Nanotechnology, National Center for Scientific Research ``Demokritos'', 15310 Athens, Greece}

\author{Iryna Omelchenko}
\affiliation{Institut f{\"u}r Theoretische Physik, Technische Universit{\"a}t Berlin, Hardenbergstra\ss{}e 36, 10623 Berlin, Germany}

\author{Eckehard Sch{\"o}ll}
\affiliation{Institut f{\"u}r Theoretische Physik, Technische Universit{\"a}t Berlin, Hardenbergstra\ss{}e 36, 10623 Berlin, Germany}

\author{Philipp H{\"o}vel} 
\affiliation{Institut f{\"u}r Theoretische Physik, Technische Universit{\"a}t Berlin, Hardenbergstra\ss{}e 36, 10623 Berlin, Germany}
\affiliation{Bernstein Center for Computational Neuroscience Berlin, Humboldt-Universit{\"a}t zu Berlin, Philippstra{\ss}e 13, 10115 Berlin, Germany}

\author{Astero Provata}
\affiliation{Institute of Nanoscience and Nanotechnology, National Center for Scientific Research ``Demokritos'', 15310 Athens, Greece}

\begin{abstract}
We study numerically the development of chimera states in networks of nonlocally coupled oscillators whose limit cycles
emerge from a Hopf 
bifurcation. This dynamical system is inspired from population dynamics and consists of three
interacting species in cyclic reactions. The complexity of the dynamics arises from the presence
of a limit cycle and four fixed points. 
When the bifurcation parameter increases away from the Hopf bifurcation
the trajectory approaches the heteroclinic invariant manifolds of the fixed points producing spikes, followed
by long resting periods. We observe chimera states in this spiking regime as a coexistence of coherence (synchronization) and incoherence (desynchronization) in a one-dimensional ring with nonlocal coupling, and demonstrate that their multiplicity depends  both on the system and the coupling parameters. 
We also show that hierarchical (fractal) coupling topologies induce traveling multichimera states.
The speed of motion of the coherent and incoherent parts along the ring is computed through the Fourier
spectra of the corresponding dynamics.
\end{abstract}


\date{\today}

\maketitle

\section{Introduction}
\label{sec1}

The recent influence of the theory of networks \cite{strogatz:1998,barabasi:2003,barabasi:2009} on the
classic field of coupled oscillatory units \cite{anishchenko:2007} 
has led  to the discovery of a plethora of novel phenomena, especially when the coupling between units
becomes more sophisticated, mimicking naturally interacting systems.
Among the complex oscillatory patterns that may emerge, are the so-called ``chimera states'' \cite{kuramoto:2002,strogatz:2004},
which have recently attracted a lot of attention. These are states, in which identically coupled units
spontaneously develop  coexisting synchronous (coherent) 
and asynchronous (incoherent) parts. 

Many recent theoretical works have focused on the study of chimera states in a variety of physical systems.
Typical models that have been numerically
investigated include the Kuramoto phase oscillator \cite{abrams:2008,laing:2012,KO08},
periodic and chaotic maps \cite{OME11,omelchenko:2012}, the Stuart-Landau model
\cite{laing:2010,ZAK14}, the Van der Pol oscillator \cite{OME15a} as well as models addressing neuron dynamics such the
FitzHugh-Nagumo oscillator \cite{omelchenko:2013}, the Hindmarsh-Rose model \cite{hizanidis:2013}, the so-called SNIPER
model of excitability type-I \cite{VUE14a}, or the Hodgkin-Huxley model \cite{SAK06a}.
Moreover, chimera states have been reported in populations of coupled pendula \cite{bountis:2014}, in autonomous
Boolean networks \cite{ROS14a},
in one-dimensional superconducting meta-materials \cite{lazarides:2015}, and time-varying networks \cite{BUS15}.

Following the theoretical predictions, chimera states were experimentally verified 
for the first time in populations of coupled chemical oscillators \cite{tinsley:2012} 
and in optical coupled-map lattices realized by liquid-crystal light modulators 
\cite{hagerstrom:2012}. Recently, in a purely mechanical experiment involving two groups 
of identical metronomes, it was shown that chimeras emerge naturally as a coexistence 
of two competing synchronization patterns \cite{martens:2013}.
Chimeras were also realized in experiments involving electronic nonlinear oscillators with delay 
\cite{LAR13} and electrochemical oscillator systems \cite{SCH14a,WIC13}.

Chimera states find increasing interest due to the possible connections
 to various phenomena observed 
in biological and social systems. 
As stated in the review paper of Panaggio and Abrams \cite{panaggio:2014}, 
chimera states could possibly explain the phenomenon of unihemispheric sleep observed 
in dolphins and some birds which sleep with half of their brain \cite{ma:2010} 
as well as ventricular fibrillation \cite{davidenko:1992}.

In the current study, we attempt to bring in evidence chimera states in the fields of 
chemical reaction kinetics and ecology. Ecological models in the form of reactive dynamical systems
have been previously used to describe the interactions between multiple
species (referred to also as particles). In most cases the dynamics follows a mean field (MF) rate equations approach, which 
describe global dynamical features 
\cite{murray:1993,may:2001}. In addition, numerical simulations are employed to address
specific details of the spatial and temporal characteristics of the systems 
\cite{llv:1999,frachebourg:1996,tsekouras:2001}.
Along the same lines, dynamical
systems describing interactions between different chemical species are studied. 
Common examples involving heterogeneous 
catalytic dynamics are the reactions $NO + CO$ \cite{imbihl:1995,imbihl2:1995}, 
$CO + O_2$ \cite{imbihl2:1995,noussiou:2007},
 $CO + OH$ \cite{anderson:1995}, all taking place on a platinum catalyst,
the  reaction   $CO + 2H_2$ on various catalysts \cite{govender:2008}, etc. All above processes are 
described as dynamical systems using the Langmuir-Hinshelwood mechanism \cite{ertl:1994}. 
Analogous modeling is extensively used for the well known 
oscillatory Belousov-Zhabotinsky reaction \cite{epstein:1998,nicolis:1977}.
In epidemiology, reactive systems are frequently used where 
the species correspond to infected individuals, susceptible or recovered ones,
and they can dynamically infect other parts of the population 
\cite{anderson:1991,delitala:2004,pastor:2001}. 
All these systems are commonly viewed as reactive dynamics 
involving reaction, birth (or adsorption), and death (desorption) processes.

In the field of experimental reactive dynamics there have already 
been several studies discussing the existence of chimera states.
Some interesting works include the photosensitive Belousov-Zhabotinsky
reaction \cite{tinsley:2012} as well as experiments with 
electrochemical oscillators \cite{SCH14a,WIC13}. Similarly,
 a social analogue of a chimera state was reported in \cite{gonzalez:2014}, 
where the authors investigated the emergence of 
localized coherence in two interacting populations of social agents.
Two social models were implemented, the Axelrod model for culture dissemination \cite{axelrod:1997} 
and the bounded confidence model by Deffuant et al. \cite{deffuant:2000}. 
Different synchronization patterns were obtained, including chimera states 
where one subpopulation of agents remained synchronized while the other was desynchronized.

It is interesting to note here that in all experiments as well as in the social systems presented above,
the subpopulations of elements are \textit{ab initio} prepared and serve as a basis for the synchronized/desynchronized
activity. On the other hand, in the numerical investigations cited above and in the current study, 
the coherent/incoherent regions emerge spontaneously
from the interplay between the system parameters, dynamics, and initial conditions.

 The main ingredients for the emergence of a chimera state in coupled oscillatory
dynamics is the stable oscillatory pattern of the single elements together with
 the nonlocal character of the interactions. 
A toy model which exhibits well defined oscillations in the form of a limit cycle
 and is relevant in ecological, chemical, and social dynamics  
is the Lattice Limit Cycle (LLC) model, proposed in \cite{llc:2002}. 
The LLC model is a nonlinear system that describes reaction, birth, and death processes and 
studies how the concentrations of the specific species change with time under the 
influence of the control parameters. 
The LLC is an ideal model to study the development of chimera states in population dynamics
for three reasons:
(a) It describes common ecological/social/chemical processes using reaction, birth, and death differential equations, 
(b) its dynamics is characterized by a limit cycle,
and
(c) when needed, it allows for direct implementation on a lattice with single species occupancy for a more realistic 
representation of the dynamics.

In this study, the capacity of the LLC model to produce chimera states
will be investigated. The chimera attributes 
will be discussed as the system approaches the critical point of the Hopf bifurcation. Furthermore, we will explore the
influence of the connection topology on the chimera features. In population
dynamics, exchange in the form of spatial diffusion between individuals residing on different parts
of the system is frequent: people move from one city to another, birds migrate, individual 
animals circulate from one herd to another,  etc. In natural and social dynamics, the
circulation of individuals has often nonlocal character and thus it is natural to investigate
phenomena like chimera states which have their origin in nonlocal interactions. In particular,
it is important to consider hierarchical connectivity in the coupling of the LLC
dynamics, since the habitats of populations
have often hierarchical/fractal distributions and morphologies: the city locations are fractally 
distributed \cite{batty:2008}, the mammal and bird habitats are located on fragmented landscapes
that affect their community dynamics \cite{sole:1999,campos:2013,buchmann:2013}, and
the viruses spread in complex socio-geographic networks \cite{wallace:1994,barthelemy:2005}.
As we will see in the sequel, hierarchical connectivity induces traveling chimera states,
a novel feature in this field.

The work is organized as follows: 
In Sec.~\ref{sec:LLC} we recapitulate the main  properties of the 
the LLC model, the MF approach,  its temporal behaviour and bifurcation
scenario. In Sec.~\ref{sec:chimeras} we define the system's
spatial geometry on a ring with classical nonlocal coupling and 
we numerically demonstrate the
emergence of chimera states in dependence of the coupling range and 
the distance from the critical
point. In Sec.~\ref{sec:chimera-fractals} we introduce fractal
connectivity matrices and give evidence of traveling chimeras. 
Additional results on connectivity with gaps are included
in the Appendix. In the concluding section
we summarize our results and discuss open problems.

\section{Parametric Study of the Lattice Limit Cycle Model}
\label{sec:LLC}
Originally, the LLC scheme \cite{llc:2002} was devised as a 
highly nonlinear model for a cyclic
reaction-diffusion process with predator-prey interactions
among three particles $X$, $Y$, and $S$. The reaction scheme reads: 
\begin{subequations}
\label{eq01}
\begin{align}
\label{eq1a} 
2X+2Y &\stackrel{p_1}\rightarrow  3Y+S \\ 
\label{eq1b}
X+S &\stackrel{p_2}\rightarrow  2X \\
\label{eq1c}
Y+S &\stackrel{p_3} \rightarrow  2S 
\end{align}
\end{subequations}
where $p_1$, $p_2$, and $p_3$ are the interaction rates.

The system~(\ref{eq01}) being an ecological model was inspired by heterogeneous catalytic reactions, 
where single particles
are deposited on the sites of a catalyst surface. These particles interact with their
neighbors or diffuse on the surface. 
The catalyst is usually represented as a regular two-dimensional (2D) lattice (square,
honeycomb, triangular), or even as a fractal lattice with impurities \cite{zhdanov:2002}. These
catalytic applications indicate the need to represent the empty lattice sites
as a virtual species, which participates in the reaction and diffusion processes as well.
The LLC model thus involves one virtual species, $S$, which represents the
empty sites and two ``normal'' species, $X$ and $Y$, which are
engaged in the three reactions.
The scheme (\ref{eq01})
describes that the interaction of two particles $X$ and two particles $Y$
turns one of the $X$-particles into $Y$ while the other $X$-particle turns into $S$ [Eq.~(\ref{eq1a})]. 
The positions (lattice sites) where the particles are located may or may not be first neighbors, 
depending on the model (see Ref.~\cite{panagakou:2013}).
Similarly, Eq.~(\ref{eq1b}) represents the birth of $X$  if another $X$ is found in an adjacent
position, while Eq.~(\ref{eq1c}) represents the
cooperative death of a particle $Y$  leaving an empty site $S$. 

This model has been extensively studied in recent years. Initially, it was implemented on a square lattice with single occupancy per lattice site
using  Kinetic Monte Carlo (KMC) simulations.
Direct KMC realizations on the 2D square lattice with nearest neighbor 
interactions produced intricate fractal patterns and local oscillations of the species concentrations \cite{llc:2002}.
Later on, long-distance diffusion was introduced as a mixing mechanism 
allowing the species to react with all particles within a specific range,
thus giving them the possibility to change their
places in the lattice at finite or infinite distances \cite{panagakou:2013}. 
The model was also studied from the viewpoint of an abstract network of phases that was shown to have 
features of a scale-free network through calculations of the degree distribution and clustering coefficient \cite{provata:2014}. 

In the MF approach, the system is described by three 4th-order nonlinear 
differential equations for the temporal total concentrations $x$, $y$, and $s$ of the respective particles $X$, $Y$, and $S$.
After applying the conservation condition $x+y+s=1$, 
the LLC system is reduced to
the following two equations:

\begin{subequations}
 \label{eq02}
\begin{align}
\frac{dx}{dt}&=-2p_1x^2y^2+p_2x(1-x-y) \label{eq2a}\\
\frac{dy}{dt}&=p_1x^2y^2-p_3y(1-x-y). \label{eq2b}
\end{align}
\end{subequations}
\begin{widetext}
The LLC system (Eq.~(\ref{eq02})) has four fixed points: one saddle-point $Q_1 = (0,0)$, 
two other fixed points $Q_2 = (0,1)$ with an unstable eigenvector direction
along the $y$-axis and $Q_3 =(1,0)$ with a stable eigenvector direction along the $x$-axis (each having one zero eigenvalue \cite{llc:2002}),
and one nontrivial fixed point whose coordinates depend on the system parameters:
$Q_4=\left(\sqrt[3]{\frac{p_3^2}{p_1p_2}\left[1+K\right]}
+\sqrt[3]{\frac{p_3^2}{p_1p_2}\left[1- K\right]}, \sqrt[3]{\frac{p_2^2}{8p_1p_3}\left[1+ K\right]}
+\sqrt[3]{\frac{p_2^2}{8p_1p_3}\left[1- K\right]}\right)$ 
with $K=\sqrt{\frac{1+\left(2p_3+p_2\right)^3}{27p_1p_2p_3}}$.
In the parameter space $(p_1,p_2)$ for fixed $p_3$, $Q_4$ is either a stable node or a stable focus which becomes unstable
through a supercritical Hopf bifurcation \cite{llc:2002}. 
Because of the physical condition $x,y,s \ge 0$ the flow is always directed to the inside of the reaction
simplex $x>0,y>0,x+y<1$ which follows from mass-action kinetics.
\end{widetext}

\begin{figure}[ht!]
\includegraphics[clip,width=\linewidth,angle=0]{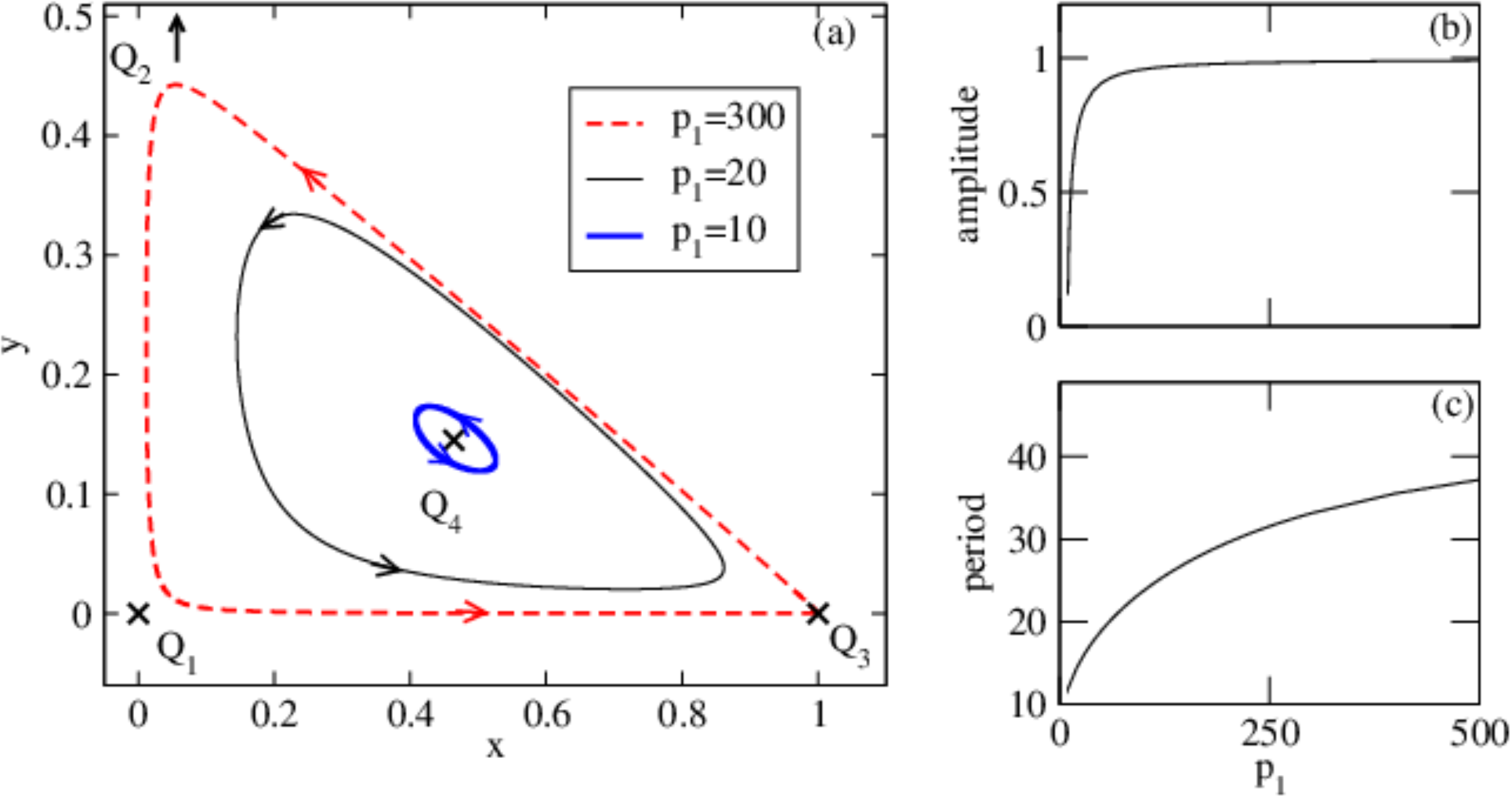}
\includegraphics[clip,width=\linewidth,angle=0]{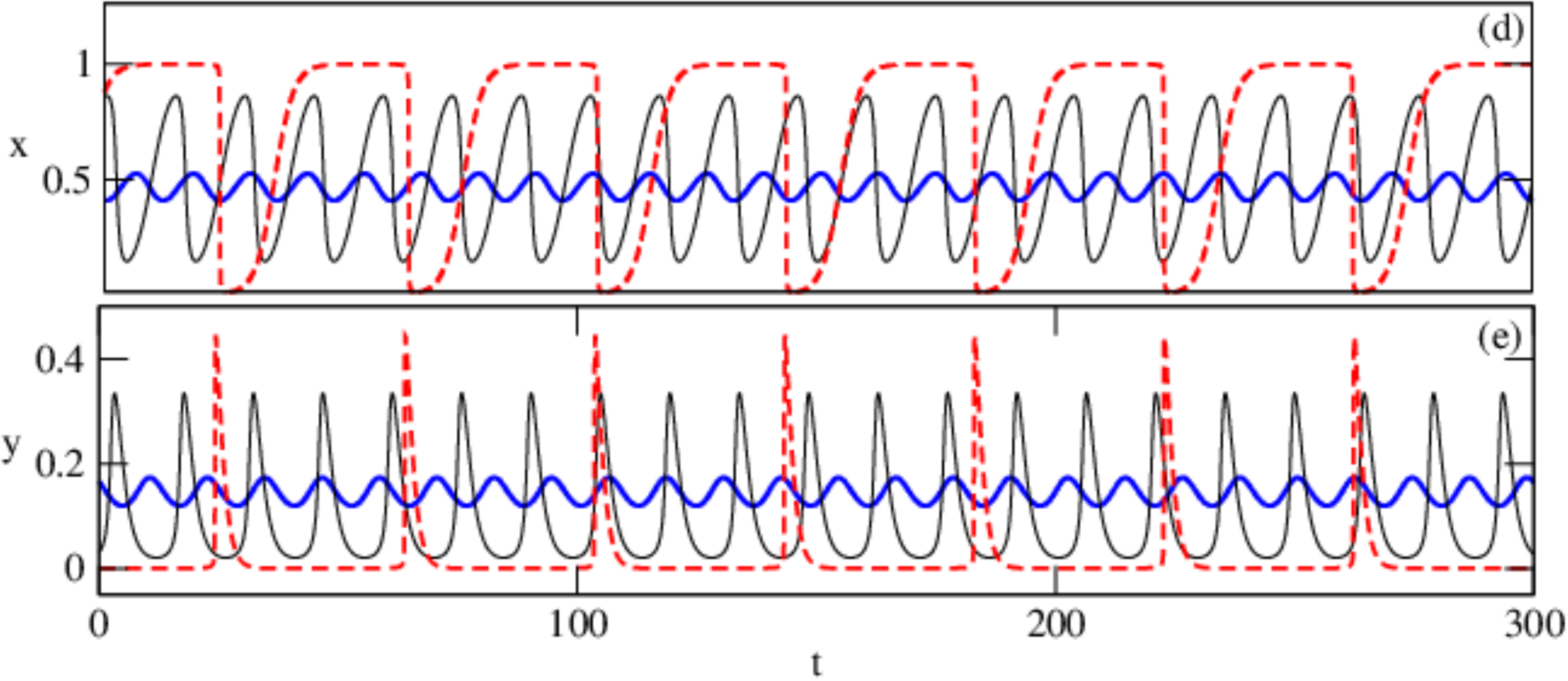}
\caption{\label{fig:01} (Color online) (a) Limit cycle for various values of $p_1$ above the Hopf bifurcation. Fixed points $Q_1$, $Q_2$ (not shown), $Q_3$, and $Q_4$ (for $p_1=10$) 
are marked with crosses. (b) Peak-to-peak amplitude of the $x$-variable
and (c) Period of the limit cycle as a function of $p_1$. (d) and (e) show the time series of $x$ and $y$, respectively,
corresponding to the limit cycles in (a). Other parameters: $p_2=0.5$ and $p_3=0.8$.
}
\end{figure}

Hereafter, we fix the parameters $p_2=0.5$ and $p_3=0.8$, while 
$p_1$ will be used as a control parameter determining the distance from the bifurcation point.
For the above values of $p_2$ and $p_3$, it can be found that the Hopf bifurcation takes place at $p_1^{crit}=9.82$. 
Figure~\ref{fig:01}(a) shows the limit cycle for three different values of $p_1>p_1^{crit}$, while the
amplitude 
and period of the limit cycle as a function of $p_1$ is plotted in Figs.~\ref{fig:01}(b) and (c), respectively.
The $x$ and $y$ time series corresponding to the limit cycles of Fig.~\ref{fig:01}(a) are depicted in
Figs.~\ref{fig:01}(d) and (e), respectively.
In the time series of the $y$-variable, in particular, the sharp spikes followed by long resting periods are visible
for large values of $p_1$. 

In the current study, we consider a setup based on MF oscillators arranged on
a cyclic one-dimensional network (ring), interacting linearly with one another. The use of the MF dynamics on all
network sites
mimics the interactions between populations, where each population is composed of a mixture of $x$, $y$, and $s$. While
the nodes interact with one another via nonlocal coupling as will be explained in the next section, the
resident population follows the MF approach on each node given by Eqs.~(\ref{eq02}). This is 
a different approach than the previous KMC
simulations \cite{tsekouras:2001,llc:2002}, where each node (lattice site) is occupied by a single particle. The properties of this particular arrangement of the populations is discussed in detail in the following section.

\section{Multichimera States in the Lattice Limit Cycle Model}
\label{sec:chimeras}
We consider $N$ nonlocally coupled LLC oscillators on a one-dimensional ring according to the following scheme:
\begin{subequations}
 \label{eq:03}
\begin{align}
\frac{dx_k}{dt}&= -2p_1x_k^2y_k^2+p_2x_k(1-x_k-y_k) +\frac{\sigma}{2R}\sum_{j=k-R}^{j=k+R}(x_j-x_k)\label{eq:03a}\\
\frac{dy_k}{dt}&= p_1x_k^2y_k^2-p_3y_k(1-x_k-y_k)+ \frac{\sigma}{2R}\sum_{j=k-R}^{j=k+R} (y_j-y_k),\label{eq:03b}
\end{align}
\end{subequations}
where the subscript $k$ refers to the node index, $k=1,\dots,N$, which has to be taken modulo $N$. Each oscillator is
coupled with its $R>0$ nearest 
neighbors on both sides with coupling strength $\sigma$. 
This introduces nonlocality in the form of a ring topology as proposed in Ref.~\cite{omelchenko:2013,OME10a}. There, the
authors considered a rotational
coupling matrix in order to achieve both direct and cross-coupling between the system variables. Our system is simpler
in the sense that it 
contains only direct coupling in the species concentrations. This reflects mobility of the species 
between nodes and no cross-interactions of species between different nodes. 

\begin{figure}[t!]
\includegraphics[clip,width=\linewidth,angle=0]{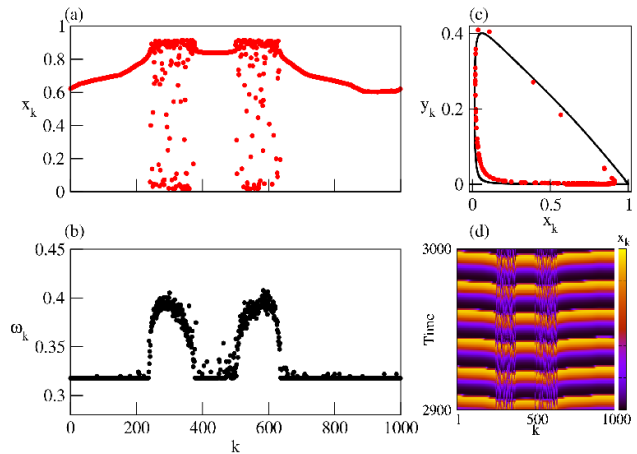}
\caption{\label{fig:02}(Color online) (a) Snapshot of the variable $x_k$ of Eqs.~(\ref{eq:03}) and 
(b) corresponding mean phase velocity
profile. 
(c) Limit cycle of the uncoupled system (black solid line) and snapshot in the $(x,y)$-plane (red dots). 
(d) Space-time plot of variable $x_k$.
Parameters: $p_1=300$, $p_2=0.5$, $p_3=0.8$, $N=1000$, $R=350$, and $\sigma =0.015$.
}
\end{figure}

A typical chimera state with two (in)coherent regions is shown in Fig.~\ref{fig:02}(a), where the $x$-concentration is
plotted as snapshot. Initial conditions in all simulations herein are taken to be randomly distributed on an ellipsoid
enclosed by the triangle with corners on the fixed points $Q_1$, $Q_2$, and $Q_3$.
As a measure indicating the existence of a chimera state we employ the {\it mean phase velocity}
 of each oscillator $\omega_k=2\pi M_k/\Delta T$, where $M_k$
is the number of periods of the $k$th oscillator during a time interval
$\Delta T$ \cite{kuramoto:2002,omelchenko:2013}. 
This quantity is depicted in Fig.~\ref{fig:02}(b) and has the typical profile:
flat, lower-valued in the coherent domains and  arc-shaped, higher-valued, in the incoherent ones. 
Figure~\ref{fig:02}(c) shows a snapshot of the $(x,y)$-plane (red dots) together with the limit cycle of the uncoupled
system
for the same parameter values. We observe that the coupling results in a limit cycle with smaller amplitude than in the
corresponding uncoupled system.
Finally, in the space-time plot of the $x$ variable in Fig.~\ref{fig:02}(d), 
it can be seen that the  chimera state is fixed in space
and has a period of approximately $18$ time units, which corresponds to the 
period of the uncoupled system for a much lower value of $p_1\approx 40$. 

The presence of chimera states in ecological and population systems can account for different 
dynamical states in different nodes of the interacting systems. For example,
in population dynamics where communities exchange individuals, the density of inhabitants
in some communities may oscillate in phase, while in other communities it may be incoherent.
Similarly, in computer networks, 
the density of users (or the data exchange) may oscillate coherently in some parts of the network, while 
in other parts it may behave asynchronously.

In the following, we discuss how the coupling range $R$ and the distance from the Hopf bifurcation, expressed 
through parameter $p_1$, influence the form and multiplicity of chimera states.

\subsection{Impact of coupling range $R$}
Chimera states with multiple domains of incoherence and coherence have been reported in several works
and are referred to as clustered chimera or multichimera states.
It is known that they may be achieved through time delay \cite{SET08} or by manipulating the range of the coupling
between oscillators \cite{omelchenko:2013,hizanidis:2013,VUE14a,MAI14}.
The range $R$ of the coupling reflects the migration range of the different species in the system.
Therefore, it is interesting to verify the formation of clustered chimera states in our system by varying this quantity.

\begin{figure}
\includegraphics[clip,width=\linewidth,angle=0]{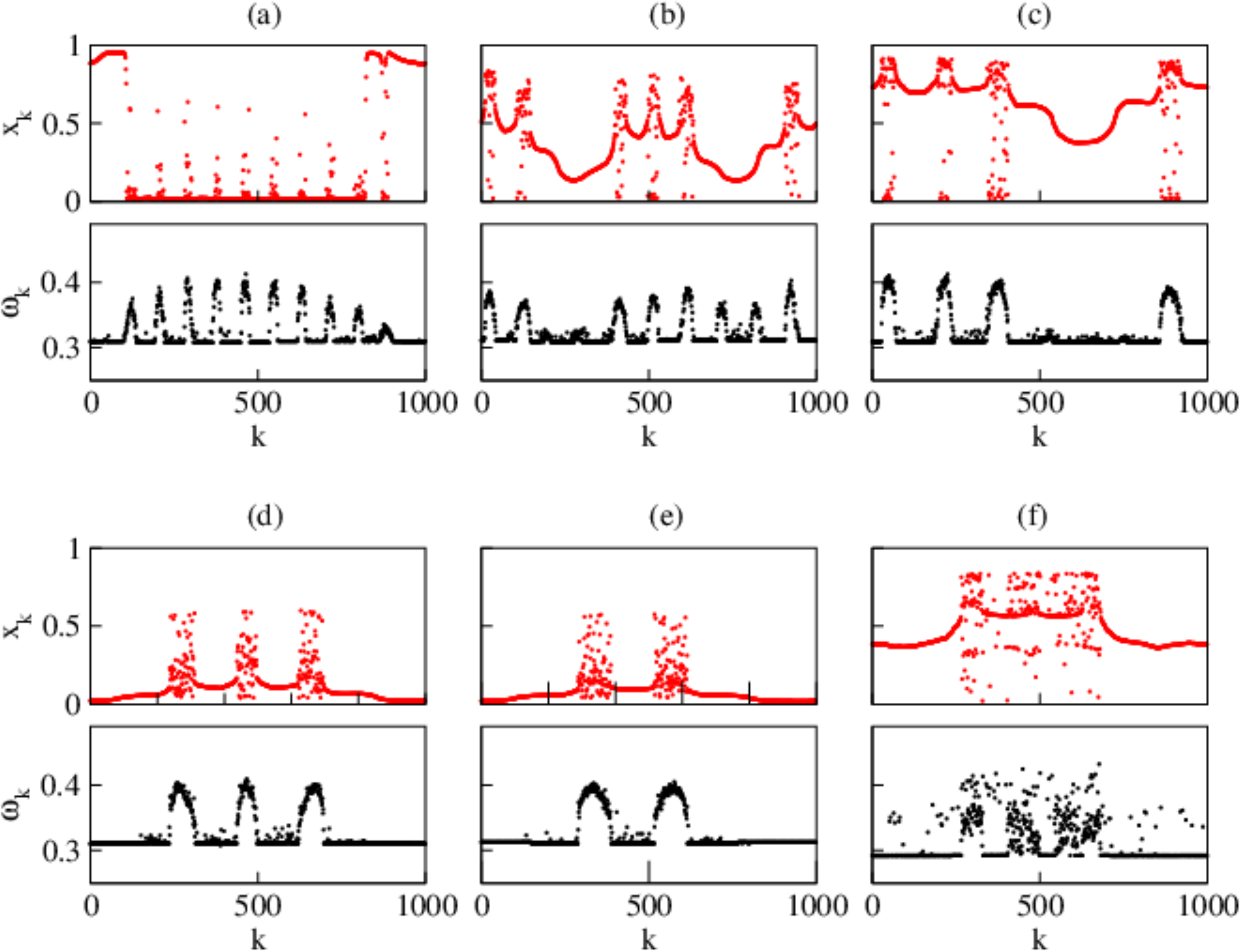}
\caption{\label{fig:03} (Color online)
Snapshots of the $x$-variable (red dots) 
and corresponding $\omega_k$ profiles (black dots) for
various values of the coupling range: (a) $R=100$, (b) $R=120$, (c) $R=190$, (d) $R=230$, 
(e) $R=280$, and (f) $R=410$. Other parameters as in Fig.~\ref{fig:02}. 
}
\end{figure}

Figure~\ref{fig:03} shows typical chimera states and corresponding mean phase velocity profiles for increasing $R$. For
low values of the coupling
range, chimera states with up to $10$ (in)coherent domains may be found [Fig.~\ref{fig:03}(a)]. For increasing $R$ the
multiplicity of the chimera state
decreases and, as $R$ approaches the value of $280$, a chimera state with $2$ (in)coherent domains
is found [Fig.~\ref{fig:03}(e)]. This $2$-chimera state persists over a large $R$-interval up to $R\approx 400$, where
it starts to deform [Fig.~\ref{fig:03}(f)].
The corresponding mean phase velocity profiles follow the multiplicity of the chimera states, exhibiting maximum values
in the incoherent domains and constant values in the coherent domains.

\subsection{Impact of bifurcation parameter $p_1$}

As mentioned in the Introduction, for fixed rates $p_2$ and $p_3$, a Hopf bifurcation occurs 
at $p_1^c\approx 9.82$ and a limit cycle starting from zero amplitude and finite period is born [see
Figs.~\ref{fig:01}(b) and (c)].
As $p_1$ increases, the limit cycle approaches the heteroclinic invariant manifolds
from $Q_2$ to $Q_1$ and from $Q_1$ to $Q_3$  
which bound the basin of attraction of the limit
cycle. The time to approach $Q_3$ becomes very large, and accordingly, its period diverges as
$p_1$ increases further.
Therefore, $p_1$ is a crucial parameter that introduces two time-scales in the system dynamics
and determines the period of the individual oscillators. Physically, $p_1$ determines the rate of reactive process 
\eqref{eq1a} and introduces a 4th-order term in the equations, since it requires simultaneous interaction of four
species (particles).
This parameter takes high values compared to the other two, $p_2$ and $p_3$, to compensate the contribution of the term
$x^2y^2$.
  
\begin{figure}[ht!]
\includegraphics[clip,width=\linewidth,angle=0]{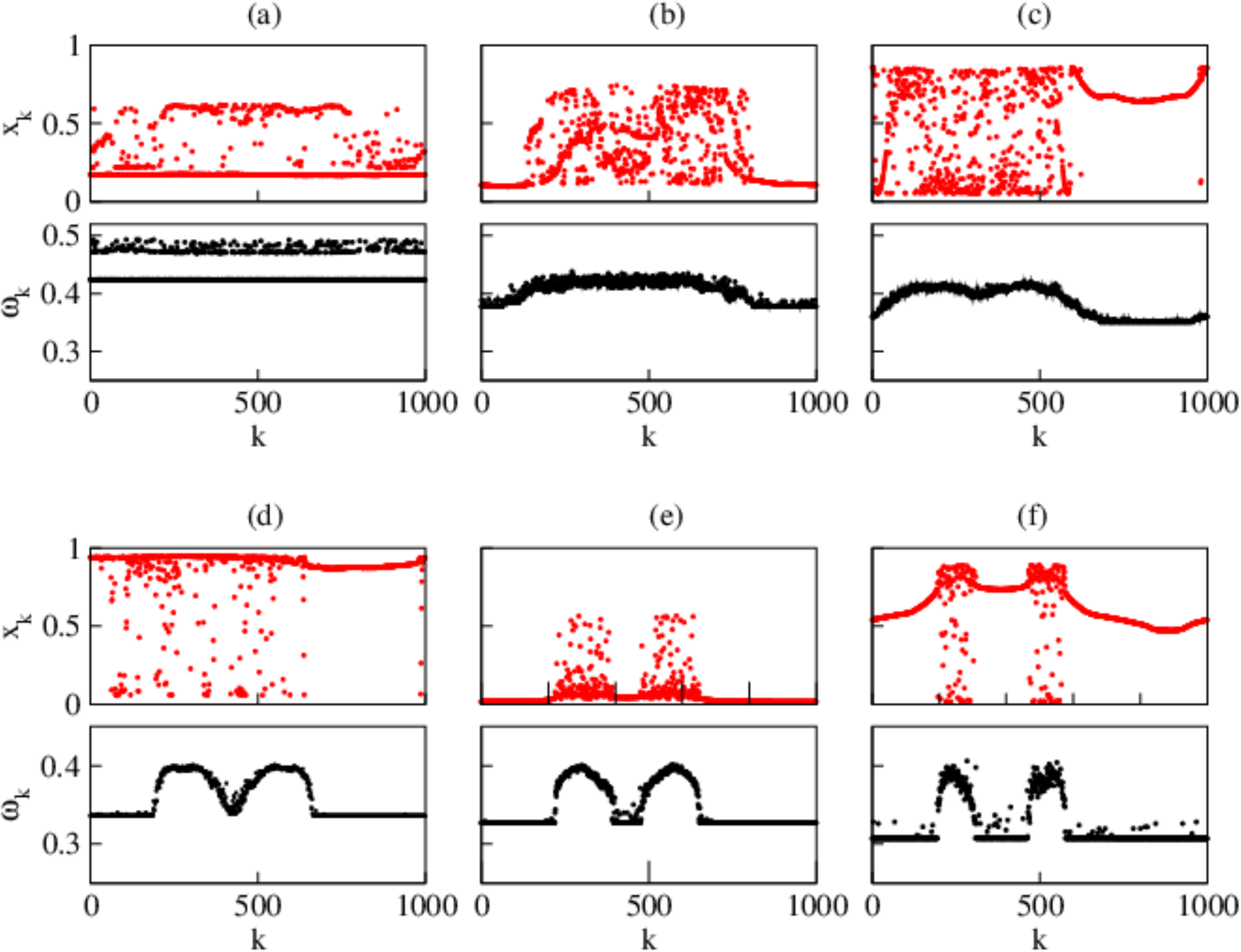}
\caption{\label{fig:04} (Color online) Snapshots of the $x$-variable (red dots) and corresponding $\omega_k$ profiles (black dots) at
time $t=3000$, for a fixed coupling range $R=350$ and various values of $p_1$: (a) $p_1=20$, (b) $p_1=40$, (c) $p_1=90$,
(d) $p_1=170$, (e) $p_1=210$ and (f) $p_1=370$.  Other parameters as in Fig.~\ref{fig:02}.}
\end{figure}

Figure~\ref{fig:04} shows typical snapshots of the $x$-variable and corresponding mean phase velocity profiles for fixed
coupling range $R=350$
and increasing $p_1$. For values of $p_1$ close to the bifurcation point [Fig.~\ref{fig:04}(a)], no chimera state is
observed but, instead, a
mixed state with no obvious spatial structure is present 
where some oscillators cluster, while others move incoherently. 
This is also reflected in the mean phase velocity profile depicted in the lower panel of Fig.~\ref{fig:04}(a), which
does not depict the characteristic arc of a chimera state.
The amplitude of the oscillations for this parameter values is small.
As we move away from the bifurcation point, the oscillators are organized in one coherent and one incoherent region,
thus, forming a $1$-chimera state [Figs.~\ref{fig:04}(b) and (c)]. 
The $\omega_k$-profile starts to obtain its typical shape, which becomes more prominent as $p_1$ attains higher values.
Then, a $2$-chimera state appears around $p_1=190$ [Figs.~\ref{fig:04}(e) and (f)]. 

A careful look at Figs.~\ref{fig:04}(c)-(f) reveals the scenario of how a chimera state with two incoherent parts is
formed. The mean phase velocity profile develops
a small dip, which increases and eventually reaches the level of the coherent part as the parameter~$p_1$ increases.
Similar scenarios have been observed for nonlocally coupled FitzHugh-Nagumo systems~\cite{omelchenko:2013}. 

\begin{figure*}[ht!]
\includegraphics[clip,width=\linewidth,angle=0]{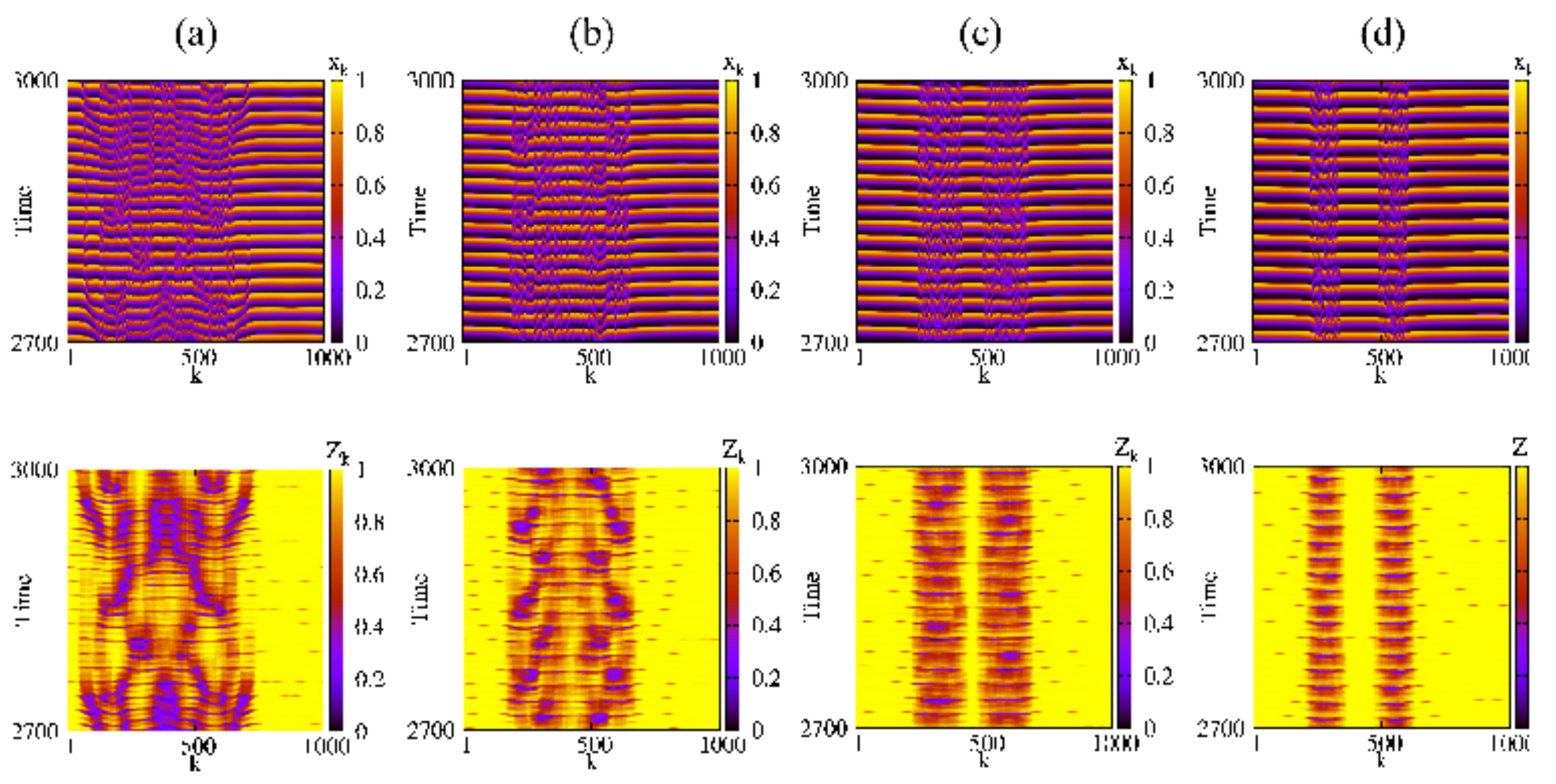}
  \caption{\label{fig:05} (Color online)
Space-time plots of the $x$-variable (top) 
and order parameter (bottom) for fixed $R=350$ and increasing $p_1$: (a) $p_1=90$, (b) $p_1=170$, (c) $p_1=210$, and (d)
$p_1=370$. 
Other parameters as in Fig.~\ref{fig:02}. 
}
\end{figure*}

The impact of $p_1$ is also depicted in Fig.~\ref{fig:05}, where the space-time plots for the variable $x$
(top panel) and the
corresponding local order parameter (lower panel) are shown for fixed $R=350$. The values of $p_1$
are chosen as $p_1=90, 170, 210,$ and $370$ in panels \ref{fig:05}(a) to \ref{fig:05}(d), respectively, which
corresponds to the snapshots in Figs.~\ref{fig:04}(c)-(f). The local order parameter is defined as follows
\cite{kuramoto:2002,OME15}: 
\begin{equation}
 Z_k=\left | \frac{1}{2\delta} \sum_{|j-k| \le \delta} e^{i\Theta_j} \right |, \quad k=1,\dots, N,
\end{equation}
where $\Theta_j=\arctan[(y_j-y_{Q_4})/(x_j-x_{Q_4})]$ denotes the geometric phase of the
$j$th LLC unit and ($x_{Q_4},y_{Q_4}$) 
are the coordinates of the nontrivial fixed point $Q_4$ of the uncoupled system. We use a spatial average with a window
size
of $\delta=25$ elements. The local order parameter $Z_k$ close to unity indicates that the $k$th unit belongs
to the coherent part of the chimera state, while $Z_k$ is less than 1 for incoherent parts.

The lower panels of Fig.~\ref{fig:05} depict the local order parameter in the time
interval $t\in[2700,3000]$, where bright (yellow) color denotes the coherent regions.
A stationary $2$-chimera state is obtained for high values of $p_1$, as shown in 
Fig.~\ref{fig:05}(d). 
As one moves closer
to the Hopf bifurcation point, the two incoherent regions merge into a large one, while the size (number of
oscillators) of the
incoherent region increases at the expense of the coherent ones [Fig.~\ref{fig:05}(a) and (b)].

\section{Traveling Chimeras for Hierarchical Connectivity}
\label{sec:chimera-fractals}

In a previous study on the FitzHugh-Nagumo model we have demonstrated that if the connectivity 
matrices have hierarchical form, the chimeras also occur as nested structures \cite{OME15}.
Here, in addition to these structures, we show evidence of a different phenomenon related to this
connectivity, namely, the occurrence of traveling coherent/incoherent regions. 

The hierarchical coupling structure involves connectivity gaps. The simplest example is a coupling matrix with two gaps.
The introduction of regular gaps induces merging and splitting of the (in)coherent parts of the chimera state
but does not produce other qualitatively different features. Representative cases are shown in the
Appendix~\ref{sec:chimera-gaps}.

\begin{figure}[ht!]
\includegraphics[clip,width=0.8\linewidth,angle=0]{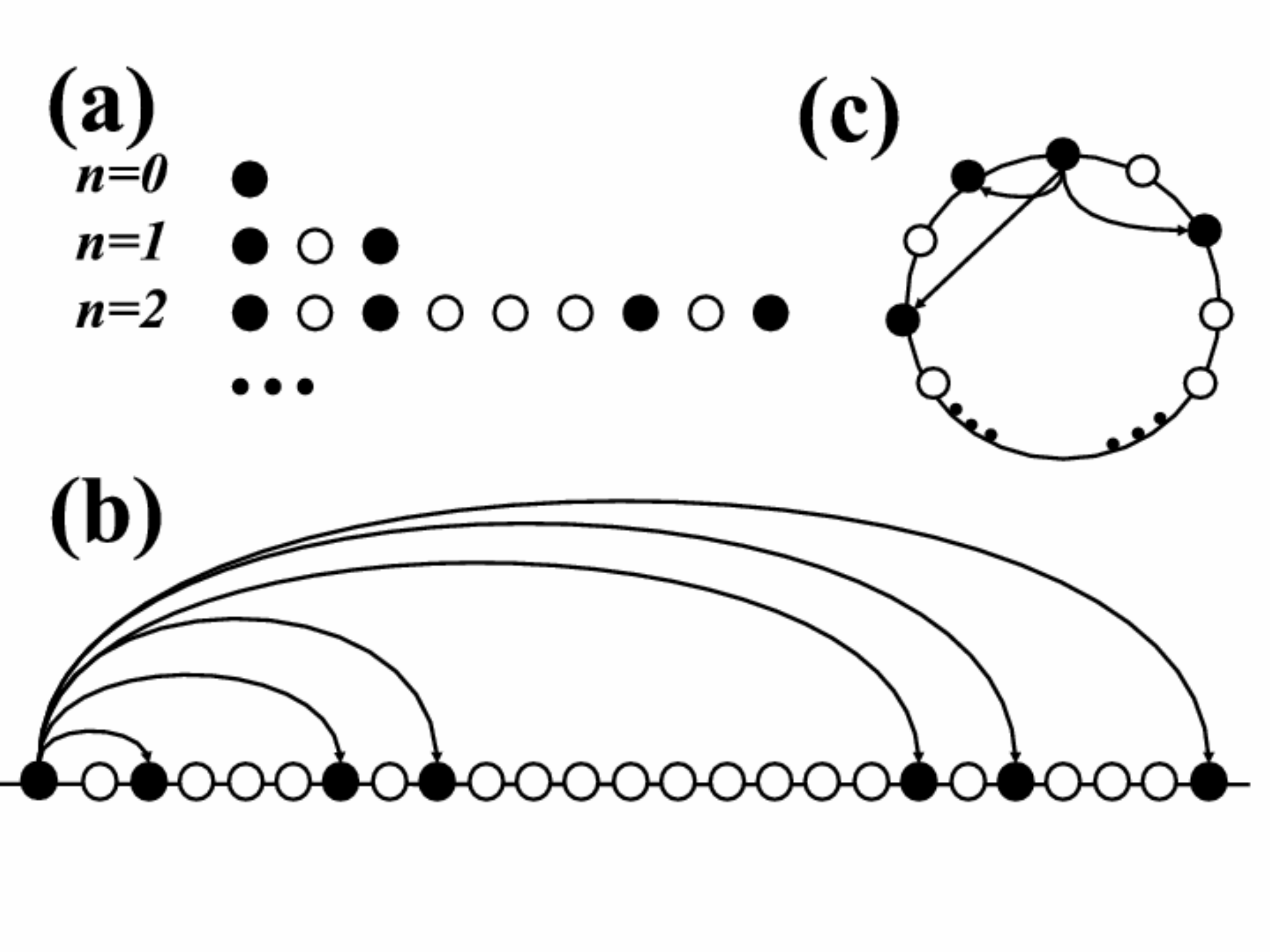}
\caption{\label{fig:06} (Color online)
Schematic representation of the 
hierarchical connectivity using the triadic Cantor set. (a) ``bottom-up'' construction of the 
triadic Cantor set, (b) linear connectivity arrangement and (c) ring connectivity 
arrangement for one reference node.
}
\end{figure}

Before we discuss the effect of the hierarchical connectivity,
we summarize the procedure to generate the corresponding network.
Figure~\ref{fig:06} illustrates this procedure in a schematic diagram \cite{feder:1988}. Let us consider a
connectivity matrix produced by an initiation string or base of size $b$, e.g. $b=3$ with a base pattern $S=$``101'' in
Fig.~\ref{fig:06}, containing $c_1$ times the symbol 1 and $c_0=b-c_1$ times the symbol 0. The specific arrangement of
the symbols on the initiation string is essential, because the hierarchical connectivity 
pattern is constructed as the $n$th iteration of the initiation string replacing 1 by the base pattern $S$ and 0 by $b$
times the symbol 0. Thus, the length of the string defining the
connectivity of the network is equal to the system size $N=b^n$ and contains $c_1^n$ times the
symbol $1$ and $N-c_1^n$ the symbol $0$. The limiting set, which is produced when the number of
iterations $n\to\infty$, is a fractal Cantor set and has fractal dimension $d_f=\ln c_1/\ln b$.
Since $d_f$ is formally defined for infinite systems only, the finite size sets used here
are called hierarchical, because they have been constructed based on a hierarchical algorithm. As a last step, we obtain
the total connectivity of the network by index shift. As we shift the iterated string along the ring,
we can determine the connectivity for each node. In this construction, each
node has precisely $c_1^n$ links to other nodes hierarchically arranged in a unique pattern and directed. The
corresponding connectivity
matrix $\left\{C^{(n)}_{kl}\right\}_{k,l=1,\dots,N}$ is given by 
\begin{align}
 C^{\{ n\} }_{kl} = \left\{
  \begin{array}{l l}
    1 & \quad \text{if both nodes $k$ and $l$ belong to the}\\
       & \quad \text{Cantor set obtained by $n$ iterations}\\ 
    0 & \quad \text{elsewhere}.
  \end{array} \right.
\label{eq4-05}
\end{align}
This connectivity matrix is of size $b^n \times b^n$ and contains a hierarchical distribution of
gaps with a variety of sizes.

In the following, we use the base size $b=6$ and $n=4$ iteration steps producing a system
of size $N=6^4=1296$. The considered initiation strings are $S=$``001111'' and $S=$``110111'', which iterated
$n=4$ times produce the connectivity patterns, consisting of 
$256$ and $625$ times the symbol $1$ and $1040$ and $671$ times the symbol $0$, respectively. 

\begin{figure}[ht!]
\includegraphics[clip,width=\linewidth]{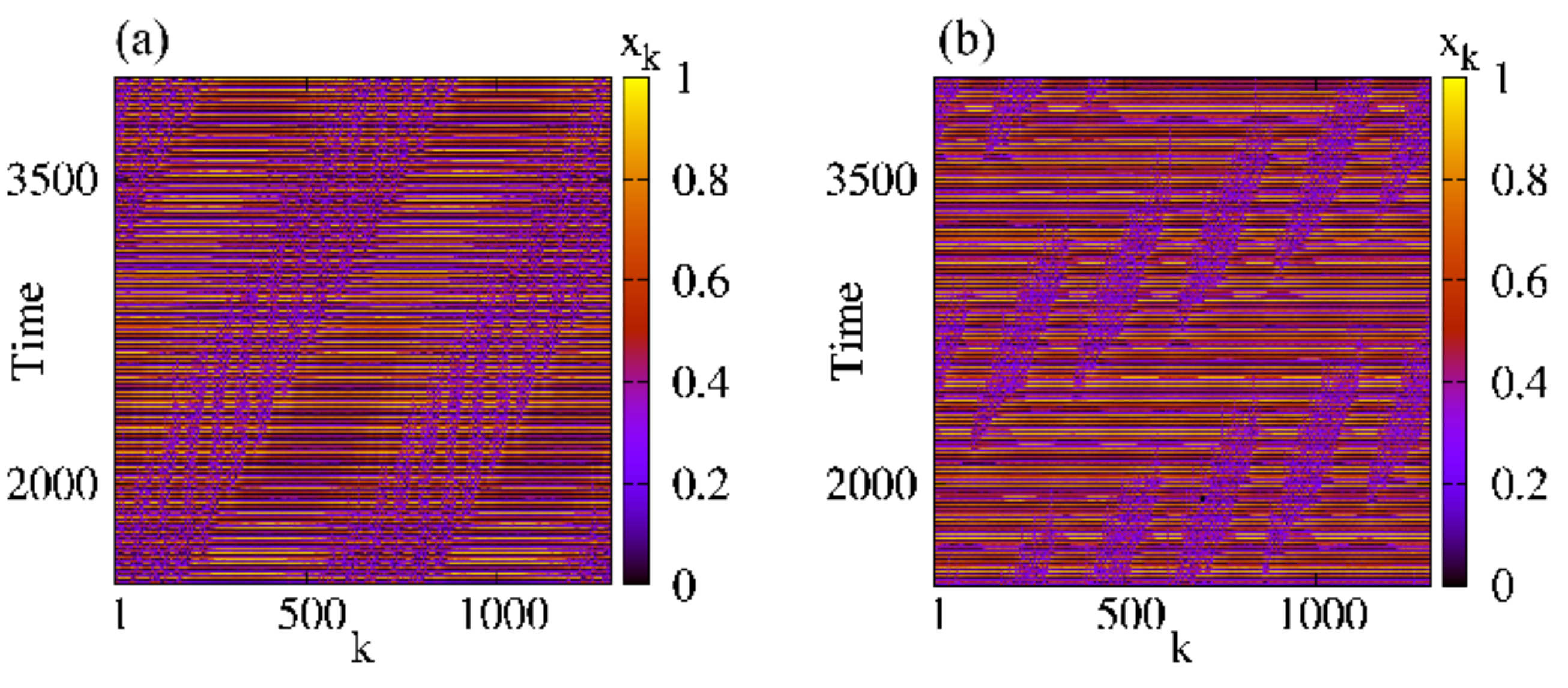}
\caption{\label{fig:07} (Color online) Space-time plots of two traveling chimeras for 
(a) $d_f=\ln 4/\ln 6=0.774$ and initiation string ``001111''
and (b) $d_f= \ln 5/\ln 6=0.898$ and initiation string ``110111''.
All other parameters as in Fig.~\ref{fig:02}.
}
\end{figure}

The motion of the (in)coherent regions in traveling chimeras are 
depicted in the space-time plots, Fig.~\ref{fig:07}(a) and (b).
From this figure we can observe a continuous translation of the (in)coherent
parts along the ring, a motion which is not observed, for example, in Fig.~\ref{fig:05},
where exemplary space-time plots of chimeras with fixed position are presented. The space-time plots 
in Fig.~\ref{fig:07} also
reveal an additional internal nested structure in both cases which are the imprints
of the hierarchical connectivity. Similar nested chimera states have been previously
demonstrated by the FitzHugh-Nagumo system without the additional feature of traveling (in)coherent parts \cite{OME15}.

\begin{figure}[ht!]
\includegraphics[clip,width=\linewidth,angle=0]{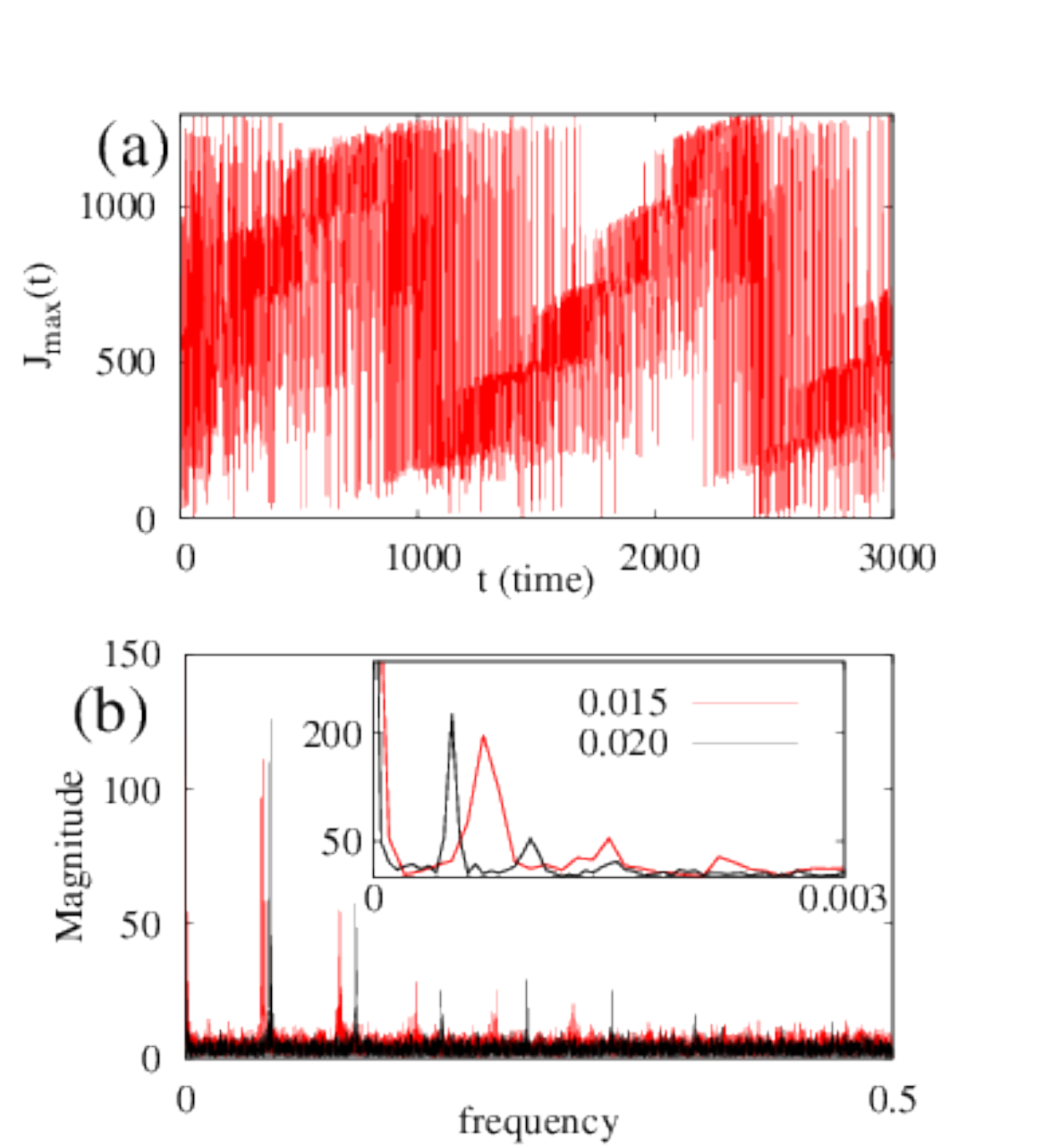}
\caption{\label{fig:08} (Color online)
(a) Position of the oscillator displaying maximum $x$-value with time for $\sigma =0.015$
 and (b)  Fourier transforms for $\sigma =0.015$
(red spectrum) and $\sigma =0.020$ (black spectrum). 
The inset in b) shows in detail the lowest part of the two spectra where the frequencies associated
with the traveling motion around the ring are discerned.
All other parameters as in Fig.~\ref{fig:07}(b).
} 
\end{figure}

In the case of traveling chimeras, the mean phase velocity is not
a good measure for coherence, because each oscillator spends part of the time 
in the coherent regions and part of the time in the incoherent ones. 
An alternative way to distinguish between stationary and
traveling chimeras is to locate the nodes attaining maximum and minimum $x$-values (or $y$-values) 
at each time step. Given that the
incoherent regions are not localized
but travel along the ring in time, we expect that the node (position on the ring) which achieves maximum
$x$-values (or $y$-values) is also subject to change. In Fig.~\ref{fig:08}(a), we record the node, 
$J_\text{max}(t)$, that exhibits maximum $x$-value at time $t$, which yields the condition:
\begin{eqnarray} 
x_{J_\text{max}}(t)=\max{ \{x_1(t), x_2(t), \cdots, x_N(t) \} }.
\label{eq46}
\end{eqnarray} 
Similarly, one can define the element $J_\text{min} (t)$, which exhibits minimum value at time $t$.
The position $J_\text{max}(t)$ (or $J_\text{min}(t)$) shows a periodicity, 
since each element successively
passes through the state of maximum amplitude. The period $T$ needed for a single element to
attain the state of maximum amplitude can be calculated
using the Fourier transform of the time series $J_\text{max}(t)$, Fig.~\ref{fig:08}(a). 
The Fourier transform, Fig. 8(b), presents two distinctive maxima.
For example, for $\sigma =0.015$ we first observe in Fig. 8(b) (red line)
one maximum of high frequency $f_\text{osc} = 0.056$, and short period $ T_\text{osc} = 1/f_\text{osc} = 17.86$ which
correspond to the average frequency of the oscillators.  By looking at the low frequencies (inset)
we observe an additional maximum at frequency $f_\text{tr} = 0.00071$
with long period $T_\text{tr} = 1408$, which is associated with the traveling
motion around the ring. Note that $T_\text{osc}$ is smaller than the period of the uncoupled oscillator $T(p_1=300)$
[cf. Fig.~\ref{fig:01}(c)], because each unit is -- from time to time -- part of the incoherent domain, which oscillates
faster than the coherent region. Furthermore, the traveling period $T_\text{tr}$ can also be inferred from the time series
$J_{\text{max}}$ shown in Fig.~\ref{fig:08}(a). The black curves in Figure~\ref{fig:08}(b) refer to a larger
coupling strength $\sigma=0.02$ than the one used in all previous plots ($\sigma=0.015$). 
Comparing the spectra for the two $\sigma$ values
in Fig.~\ref{fig:08}(b) we find that the frequency $f_{tr}$ moves
to lower values as $\sigma$ increases.

\begin{figure}[ht!]
\includegraphics[clip,width=0.6\linewidth,angle=270]{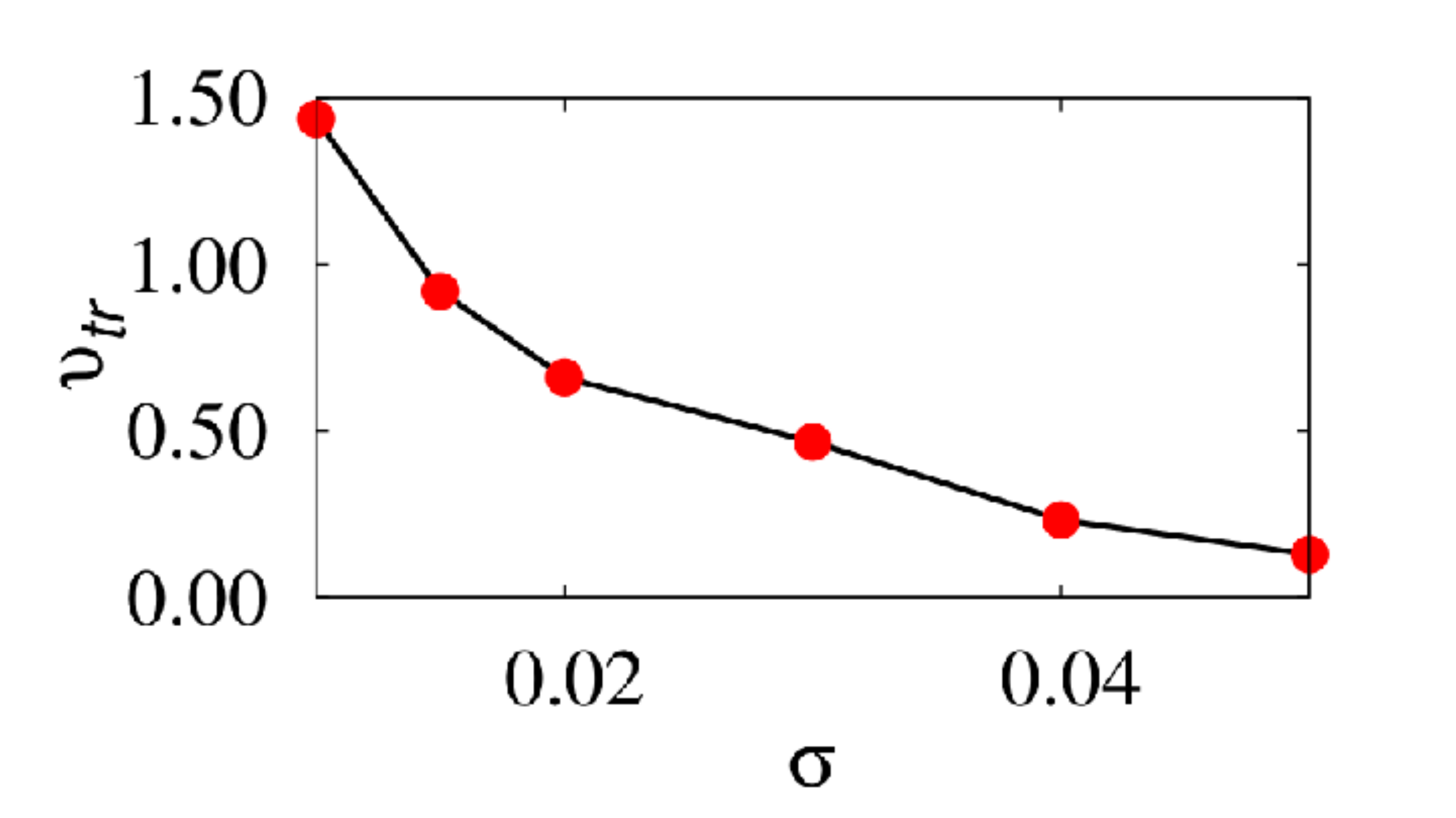}
\caption{\label{fig:09} (Color online)
Traveling speed $v_\text{tr}$ of (in)coherent regions around the ring.
All other parameters as in Fig.~\ref{fig:07}(b).
}
\end{figure}

This is further detailed in Fig.~\ref{fig:09} that shows the traveling speed $v_\text{tr}$ of the domains. 
During one period $T_\text{tr}$, the coherent
and incoherent regions cover the entire ring consisting of $N$ elements, thus,
\begin{eqnarray} 
v_\text{tr}=N/T_\text{tr}=N f_\text{tr}
\label{eq47}
\end{eqnarray}
The chimera traveling speed $v_\text{tr}$ decreases for increasing coupling strength $\sigma$, i.e. as the coupling
constant
becomes larger, the incoherent regions slow down around the ring. We recall here the results of Sec.~\ref{sec:chimeras}
where   
chimera states are only observed for low coupling strengths in the LLC model.

As alternative measure for the calculation of the traveling speed we may use the spectrum
of single oscillators (not shown). Since each oscillator belongs periodically either to the coherent or the incoherent
regions, it will participate once during each period in the regions which demonstrate
maximum (or minimum) amplitude. Then by calculating the spectrum of a single oscillator and selecting
the lowest frequency we obtain an estimation of the chimera traveling speed $v_\text{tr}$, based on
one oscillator. Average over the $N$ elements needs to be taken to represent 
the overall chimera speed $v_\text{tr}$. 

\section{Conclusions}
\label{conclusions}
In the present work, we have demonstrated for the first time the existence of chimera states in a population dynamics
model, reflecting
the various tendencies of local communities to behave (oscillate) coherently or incoherently within the same
network. 
We have shown that the dynamics of the nonlocally coupled system strongly depends on the character of the local behavior
of the nodes. When the parameters of the individual 
LLC nodes are close to the Hopf bifurcation, chimera states cannot be observed in the system. In this regime,
all strongly nonlinear 4th-order terms become weak and the corresponding limit cycle oscillations lose their spiking
form 
resembling simple harmonic motion of small amplitude.
When the internal bifurcation parameter of each node increases away from the Hopf bifurcation,
the trajectories approach the heteroclinic invariant manifolds of the saddle points producing spikes followed
by long resting periods. This type of local dynamics allows for the existence of a variety of chimera and multichimera
states, which depend both on the parameters of the individual system, and the coupling.

While the introduction of single gaps in the connectivity matrix induces merging and splitting 
of the (in)coherent parts of the chimera states (see Appendix~\ref{sec:chimera-gaps}), hierarchical arrangement of gaps
results in 
the emergence of nontrivial phenomena in which the (in)coherent regions show nested structures and
travel along the ring, keeping their profiles statistically stable in time. 
Moreover, we have found that the speed of this motion decreases with increasing coupling strength. 
Complex nested chimera structures, when regarded from the viewpoint of population dynamics, show
the rich organization which can emerge in communities of nonlinearly interacting populations.

\section{Acknowledgments}
This work was supported by the German Academic Exchange Service (DAAD) and the Greek State Scholarship Foundation IKY
within the PPP-IKYDA framework. 
JH and AP acknowledge support by YDISE project within GSRT's KRIPIS action, funded by Greece and the European Regional
Development Fund of the European Union under  NSRF 2007-2013 and the Regional Operational Program of Attica. PH
acknowledge support by BMBF (grant no. 01Q1001B) in the framework of BCCN  Berlin (Project A13). IO, ES, and PH
acknowledge support by DFG in the framework of the Collaborative Research Center 910.  The research work was partially
supported by the European Union's Seventh Framework Program (FP7-REGPOT-2012-2013-1) under grant agreement n316165.

\clearpage

\appendix

\section{Chimera states for  connectivity matrices with gaps}
\label{sec:chimera-gaps}
Previous studies have demonstrated that the introduction of gaps
in a system of coupled elements modifies the chimera states,
producing different coherence-incoherence patterns \cite{OME15}. We test this hypothesis in the case of the LLC model
and investigate whether
this effect is generic or  exclusive to specific models.
For this we consider an adjacency matrix, $\left\{C_{kl}\right\}$, $k,l=1,\dots,N$, with two
gaps, one to the left and one to the right of the node $k$, around which the connectivity is described,
called {\it reference node}. Generally,
 the size of the gaps 
and the sizes of the linked regions can vary independently
on the left and on the right of the reference node. The form of matrix element
$C_{kl}$ is:
\begin{align}
 C_{kl} = \left\{
  \begin{array}{l l}
    1 & \quad \text{if $k-R_3<l<k+R_1$}\\
           & \quad \text{ or  $k+R_1+G_R<l<k+R_1+G_R+R_2$}\\
           & \quad \text{ or  $k-R_3-G_L-R_4<l<k-R_3-G_L$}\\
    0 & \quad \text{elsewhere},
  \end{array} \right.
\label{eq4-02}
\end{align}
\noindent
where all indices are taken modulo $N$. This general form indicates that the node $k$ is linked to two
groups of nodes to the right with sizes $R_1$ and $R_2$ which are separated by a gap of size $G_R$.
A similar connectivity arrangement
with linked groups of sizes $R_3$ and $R_4$ and gap size $G_L$ applies to the left of each node. 
Symmetric connectivity is realized as a particular case of Eq.~(\ref{eq4-02}) with
$R_3=R_1$, $G_L=G_R$ and $R_4=R_2$.
The connectivity matrix Eq.~(\ref{eq4-02}), 
can be generalized further, to present multiple gaps to the left and to the right interrupting the
linked elements. An asymmetric connectivity matrix, for instance, describes a directed network.

We now test the influence of the gap size using asymmetric connectivity matrices with 
links extending only to the right of each node. Because symmetry in the connectivity 
does not influence the relative fraction of nodes belonging to the (in)coherent part
of the chimera, we only consider links to the right part of each node.
This reduction is introduced to
keep the number of parameters minimal; by using $R_3=R_4=G_L=0$ we only vary three
parameters related to connectivity,
$R_1$, $G_R$ and $R_2$.

\begin{figure}
\includegraphics[clip,width=\linewidth,angle=270]{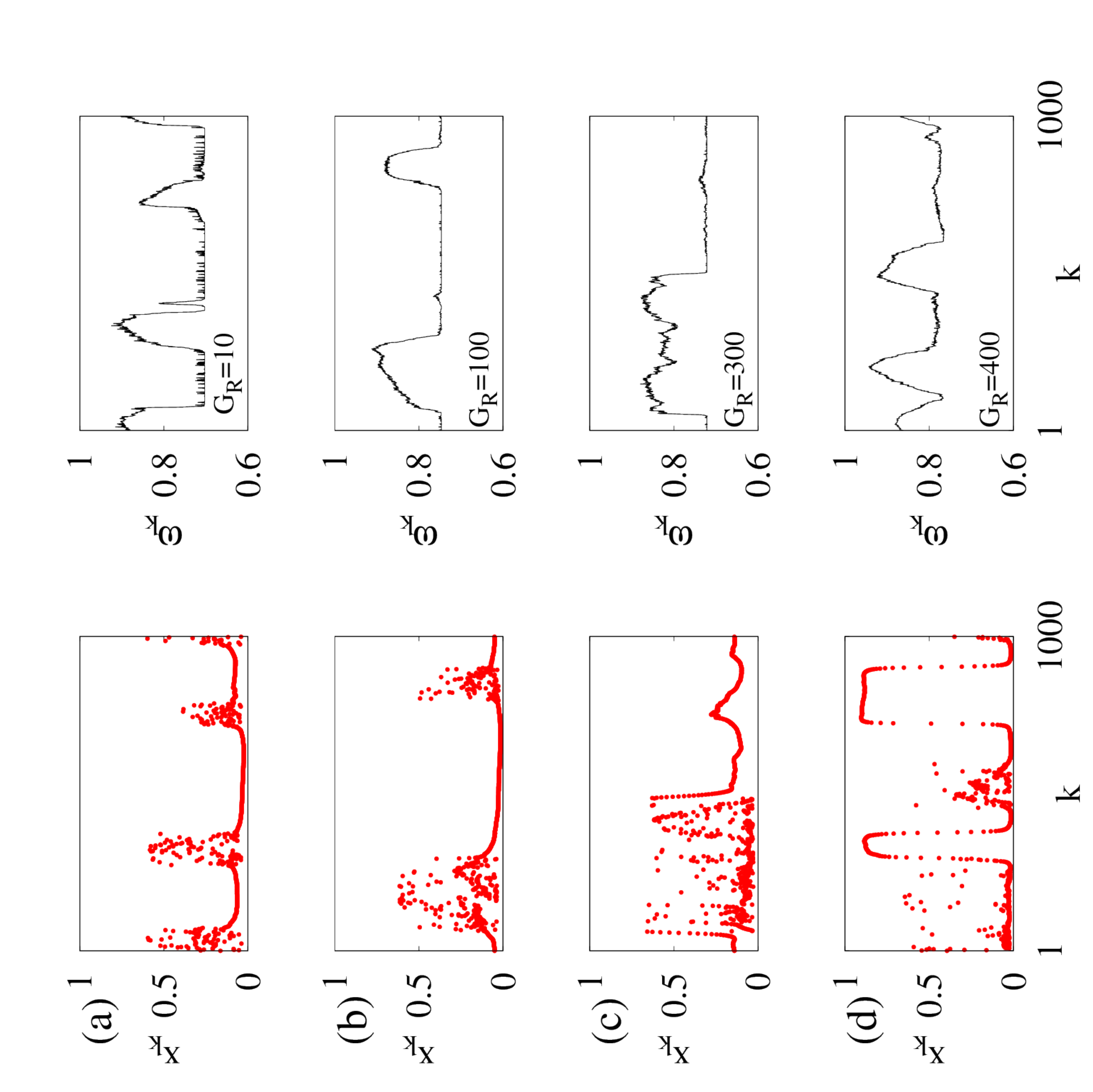}
\caption{\label{fig:app-01} (Color online)
Chimera states (left) and corresponding  mean phase velocity profiles (right) for
gap size variation. The gap sizes are indicated in the right panels, while $R_1=100$ and $R_2=300$.
All simulations start from the same initial conditions. Other parameters as in Fig.~\ref{fig:02}.
}
\end{figure}

To investigate the influence of the gap size on coherence, we keep the sizes of the two linked regions
$R_1$ and $R_2$ constant, and vary only the gap size $G_R$ in the interval $0 \le G_R\le R_1+R_2$.
In Fig.~\ref{fig:app-01} we plot the typical chimera profiles
as we increase the gap size $G_R$. The system size is kept to $N=1000$ oscillators, with parameter
values $p_1=300, p_2=0.5,p_3=0.8$, $\sigma =0.015$. The total number of connections is also kept
fixed, $R=400$, and is divided into two regions $R_1=100$
and $R_2=300$.

In Fig.~\ref{fig:app-01} we observe changes in the chimera multiplicity, ranging from one to three
(in)coherent regions, merging and splitting of coherent and incoherent regions and 
shifting of their position in space.
The change of the chimera profiles can be quantified statistically by calculating 
the following measures of coherence \cite{OME15}: 
a) the average mean phase velocity of the coherent parts, $\langle \omega_{\text{coh}}\rangle$,
b) the maximum difference of the mean phase velocities $\Delta\omega=\omega_{max}-\omega_{min}$,
c) the fraction of oscillators belonging to the incoherent parts, $N_{\text{incoh}}$, and
d) the extensive measure of incoherence $M_{\text{incoh}}$.
The relative size $N_{\text{incoh}}$
of the incoherent
parts of the chimera state is calculated as:
\begin{eqnarray}
 N_{\text{incoh}} = \frac{1}{N}\sum_{k=1}^N\Theta (\omega_k-\langle\omega_{\text{coh}}\rangle -c)
\label{eq4-03}
\end{eqnarray}
where $\Theta$ is the step function which takes the value 1 when its argument takes positive
values and zero otherwise. $c$ is a small tolerance, which in this case we set to 0.05. 
We also define the extensive, cumulative size $M_{\text{incoh}}$ of the incoherent parts as:
 \begin{eqnarray}
 M_{\text{incoh}} = \sum_{k=1}^N |(\omega_k-\langle \omega_{\text{coh}} \rangle )|
\label{eq4-04}
\end{eqnarray}
This is an extensive measure which represents the area below the arcs in the  
mean phase velocity profiles. 
$M_{\text{incoh}}$ is a measure of incoherence and is equal to zero for
fully coherent states.

\begin{figure}
\includegraphics[clip,width=0.5\linewidth,angle=270]{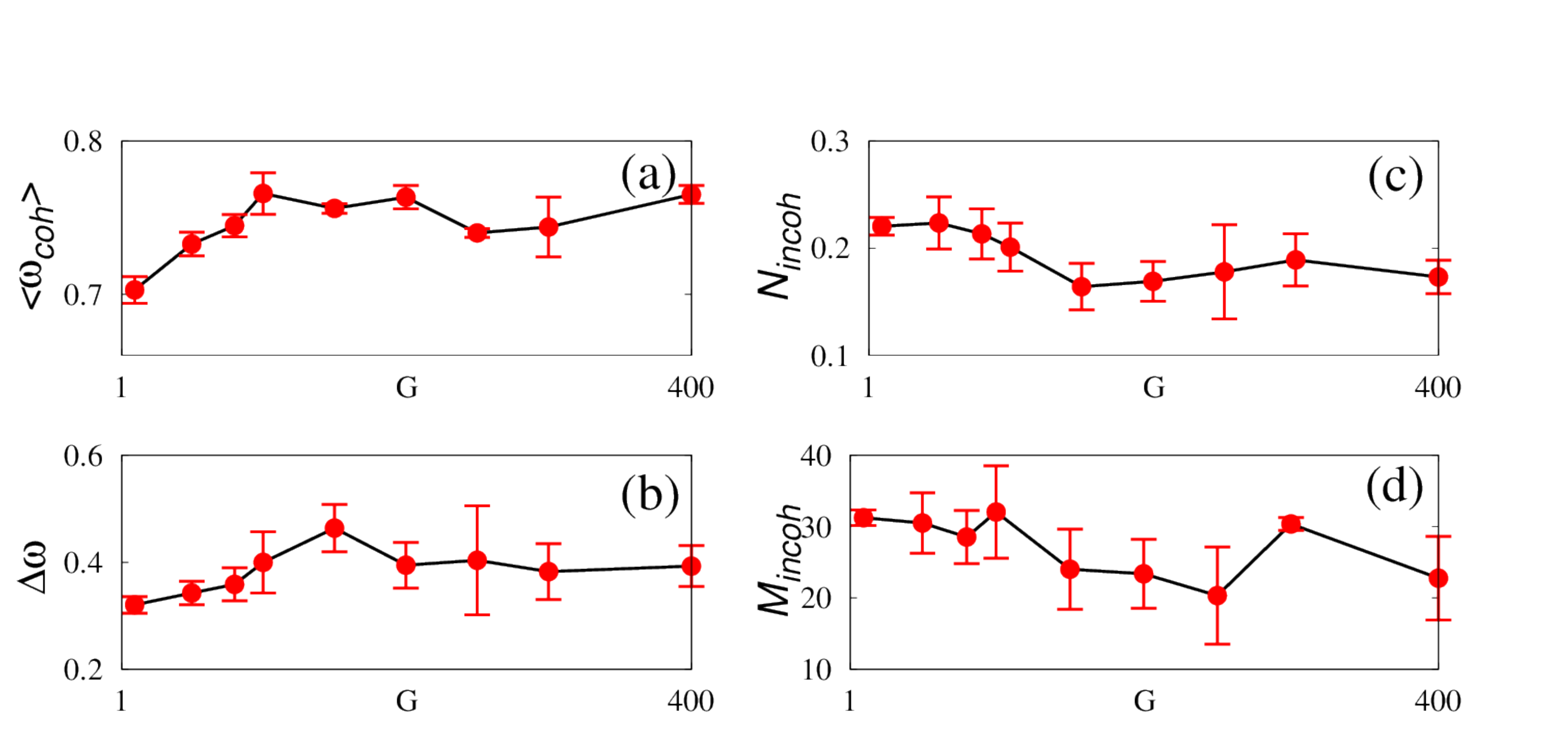}
\caption{\label{fig:app-02} (Color online)
The four measures of coherence as a function of the  gap size
$G$:
a)  $\langle \omega_{\text{coh}}\rangle $,
b)  $\Delta\omega=\omega_{max}-\omega_{min}$,
c)  $N_{\text{incoh}}$,
d)  $M_{\text{incoh}}$. Averages are
taken over 10 different sets of initial conditions. Parameter values as in Fig.~\ref{fig:app-01}.
}
\end{figure}

In Fig.~\ref{fig:app-02}  the four coherence measures a plotted, averaged over 10 different 
initial conditions. 
Variation of the gap size shows a slight increase in the $\omega_{coh}$ value for small values
of $G$, which soon attains a constant value, independent of the gap size. The other three measures
of coherence do not show any appreciable change with variations of the gap size.
Especially the number of
incoherent and coherent oscillators $N_{\text{incoh}}$ and $N-N_{\text{incoh}}$ do not change drastically
with $G$, although the number of coherent and incoherent regions may change, as indicated by
Fig.~\ref{fig:app-01}.
This means that when the gap size changes there is a redistribution of the incoherent
oscillators in one large region (Fig.~\ref{fig:app-01}(c)) or in two or three smaller ones
(Fig.~\ref{fig:app-01}(a),(b),(d)), 
so that the total number
of incoherent oscillators is conserved (see Fig.~\ref{fig:app-02}).

\begin{figure}[ht!]
\includegraphics[clip,width=\linewidth,angle=270]{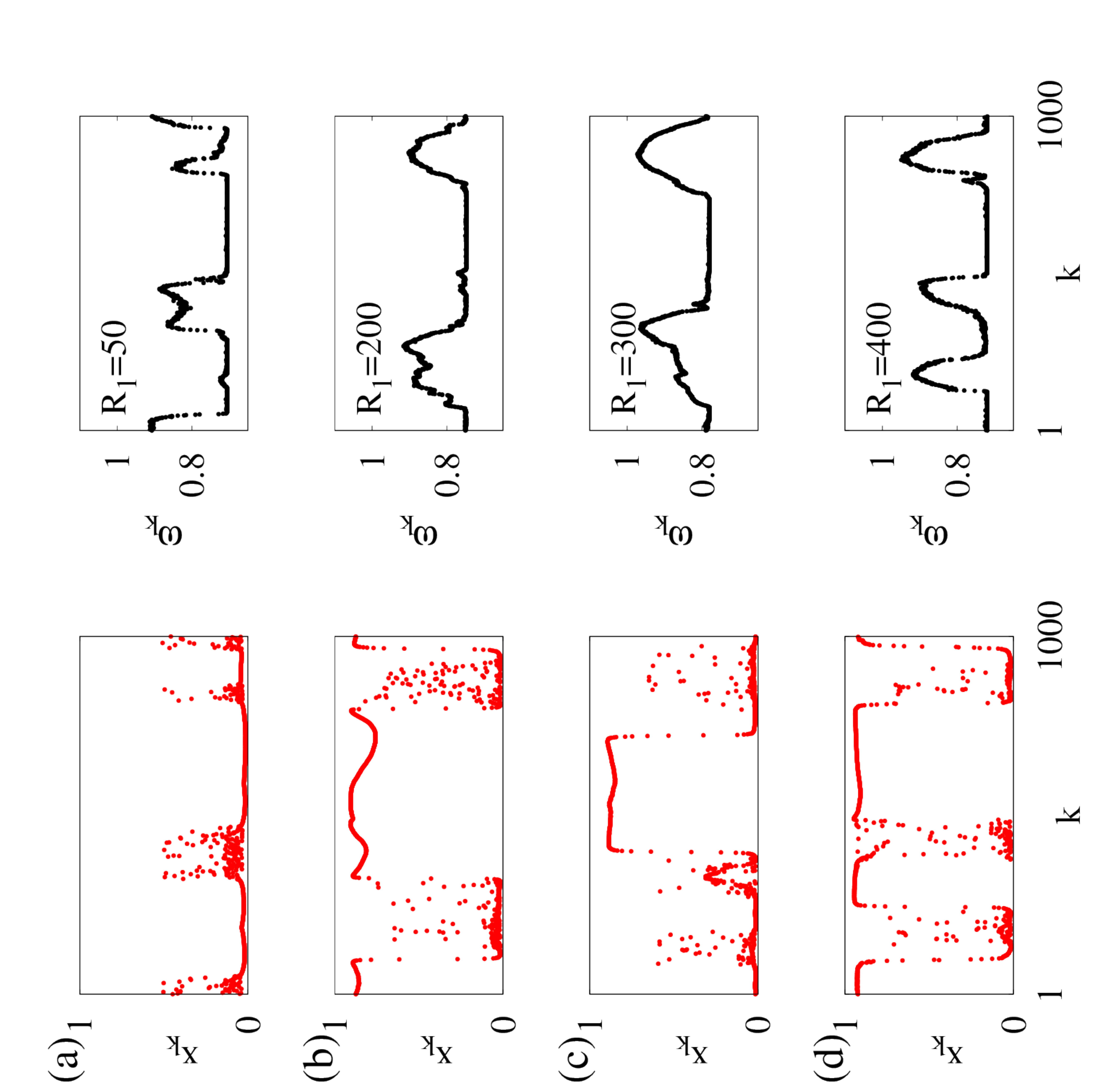}
\caption{\label{fig:app-03} (Color online)
Chimera states and corresponding  mean phase velocity with
variations on the position of a gap of constant size $G_R=100$. 
(a)  $R_1=50, R_2=350$,
(b)  $R_1=200, R_2=200$, (c)  $R_1=300, R_2=100$, and d) $R_1=400, R_2=0$.
All runs start from the same initial conditions. Other parameters as in Fig.~\ref{fig:02}.
}
\end{figure}

Earlier studies for connectivity matrices with gaps in the FitzHugh-Nagumo system have
demonstrated that the position of the gap is important and that it affects the chimera properties
modifying all four measures of coherence \cite{OME15}. 
To test this for the LLC model we produce the snapshots
and corresponding mean phase velocity profiles varying the 
sizes of the connectivity regions $R_1$ and $R_2$ but 
keeping the total number of links $R_1+R_2=400$ fixed,
while introducing a gap of constant 
size $G_R=G=100$ at various positions between the links. The results are depicted in Figure \ref{fig:app-03}.

\begin{figure}[ht!]
\includegraphics[clip,width=0.5\linewidth,angle=270]{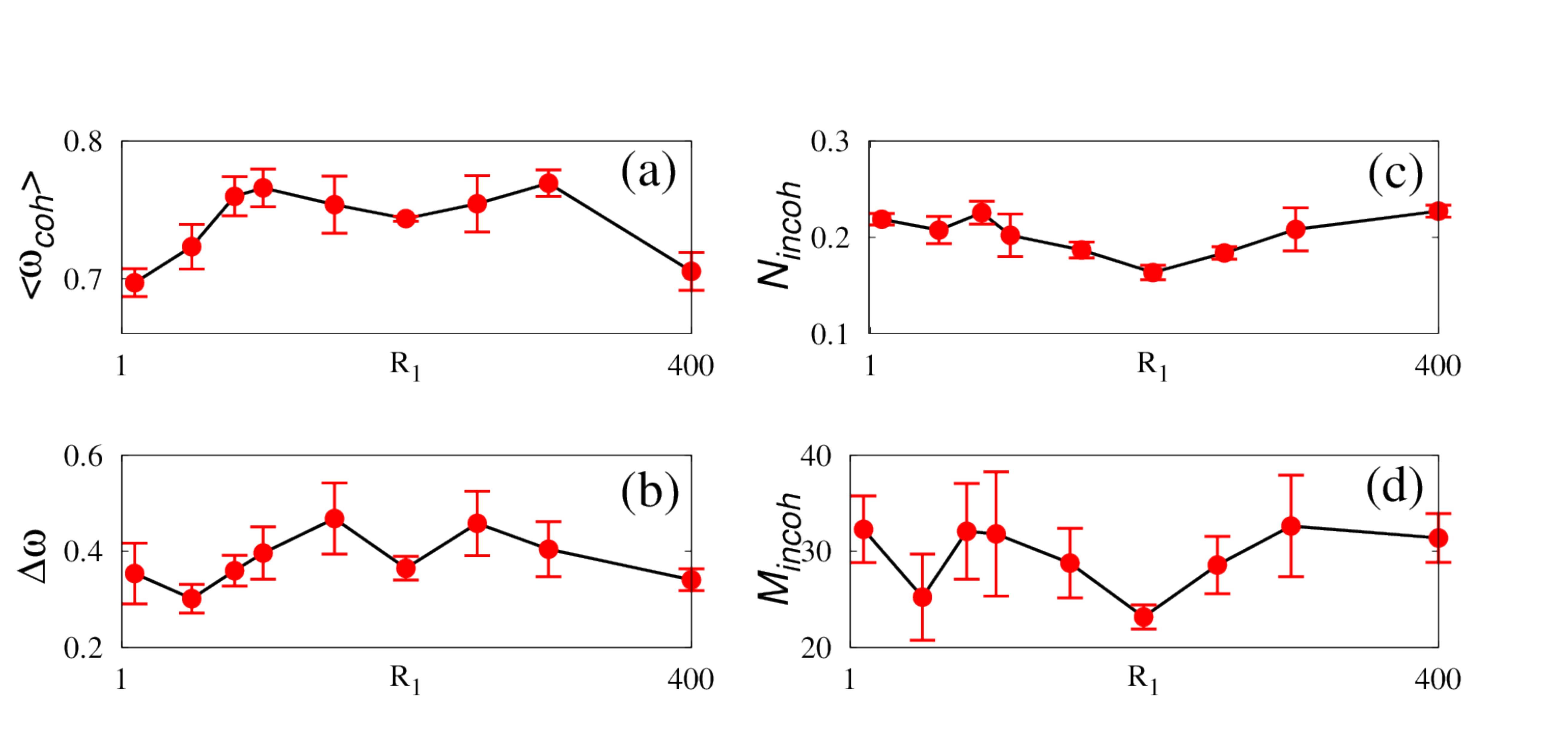}
\caption{\label{fig:app-04} (Color online)
The four measures of coherence as a function of the  
inner connectivity radius $R_1$, while $R_1+R_2=400$:
(a)  $\langle \omega_{\text{coh}}\rangle $,
(b)  $\Delta\omega=\omega_{max}-\omega_{min}$,
(c)  $N_{\text{incoh}}$,
(d)  $M_{\text{incoh}}$. The gap size is kept to $G_R=100$.
Other parameter values as in Fig.~\ref{fig:app-01}. Averages are calculated over 10 different
initial conditions. 
}
\end{figure}

In contrast to the FitzHugh Nagumo system, here we do not observe a systematic dependence
in the number of the (in)coherent regions as the gap of constant size
moves away from the reference node.
To further verify this, we plot in  Fig.~\ref{fig:app-04} the four measures of coherence
for different values of $R_1$. Averages are taken over 10 different
 different initial conditions.  All four measures seem to be constant
(up to fluctuations) and they do not show any systematic change as a function of $R_1$.


\begin{thebibliography}{10}

\bibitem{strogatz:1998}
D. J. Watts and S. H. Strogatz,
 Nature \textbf{393}, 440 (1998).

\bibitem{barabasi:2003} E. Ravasz and A.-L. Barab\'asi,
Phys. Rev. E \textbf{67}, 026112 (2003).

\bibitem{barabasi:2009} A.-L. Barab\'asi,
Science \textbf{325}, 412 (2009).

\bibitem{anishchenko:2007} V. S. Anishchenko, V. Astakhov, A. Neiman, 
T. Vadivasova, and L. Schimansky-Geier,
\textit{Nonlinear Dynamics of Chaotic and Stochastic Systems},
Springer-Verlang, Berlin 2007.

\bibitem{kuramoto:2002}  Y. Kuramoto and D. Battogtokh, 
Nonlinear Phenomena in Complex Systems \textbf{5}, 380  (2002).


\bibitem{strogatz:2004} D. M. Abrams and S. H. Strogatz, 
Phys. Rev. Lett. \textbf{93}, 174102 (2004).


\bibitem{abrams:2008} D. M. Abrams, R. Mirollo, S. H. Strogatz, and D. A. Willey, 
Phys. Rev. Lett. \textbf {101}, 084103 (2008).

\bibitem{KO08}
T.~W. Ko and G.~B. Ermentrout, Phys. Rev.~E {\bf 78}  016203  (2008).

\bibitem{laing:2012} C. R. Laing, K. Rajendran, and I. G. Kevrekidis, 
Chaos \textbf {22}, 013132, (2012).



\bibitem{OME11}
I. Omelchenko, Y.~L. Maistrenko, P. H{\"o}vel, and E. Sch{\"o}ll, Phys. Rev.
Lett. {\bf 106},  234102  (2011).
  
\bibitem{omelchenko:2012} I. Omelchenko, B. Riemenschneider, P. H\"ovel,
Y. Maistrenko, and E. Sch\"oll,
Phys. Rev. E \textbf{85}, 026212 (2012).

\bibitem{OME15a}
I. Omelchenko, A. Zakharova, P. H{\"o}vel, J. Siebert, and E. Sch{\"o}ll, (2015) arXiv:1503.03377.

\bibitem{laing:2010} C. R. Laing, 
Phys. Rev.~E \textbf {81}, 066221 (2010).

\bibitem{ZAK14}
A. Zakharova, M. Kapeller, and E. Sch{\"o}ll, Phys.~Rev.~Lett. {\bf 112},
154101  (2014).

\bibitem{omelchenko:2013} I. Omelchenko, O. Omel'chenko, P. H\"ovel,
and E. Sch\"oll,
Phys. Rev. Lett. \textbf{110}, 224101 (2013).



\bibitem{hizanidis:2013} J. Hizanidis, V. Kanas, A. Bezerianos, and T. Bountis, 
Int. J. of Bifurc. and Chaos \textbf {24}, 1450030 (2013).

\bibitem{VUE14a}
A. V{\"u}llings, J. Hizanidis, I. Omelchenko, and P. H{\"o}vel, New J.~Phys.
  {\bf 16},  123039  (2014).
  
\bibitem{SAK06a}
H. Sakaguchi, Phys. Rev.~E {\bf 73},  031907  (2006).

\bibitem{bountis:2014}
T. Bountis, V. Kanas, J. Hizanidis, and A. Bezerianos, Eur. Phys.~J. Special
Topic {\bf 223},  721  (2014).

\bibitem{ROS14a}
D.~P. Rosin, D. Rontani, N.~D. Haynes, E. Sch{\"o}ll, and D.~J. Gauthier,
  Phys. Rev.~E {\bf 90},  030902(R)  (2014).

\bibitem{lazarides:2015}
N. Lazarides, G. Neofotistos, and G. P. Tsironis, Phys. Rev.~B {\bf 91}, 054303 (2015).  

\bibitem{BUS15}
A. Buscarino, M. Frasca, L.~V. Gambuzza, and P. H{\"o}vel, Phys. Rev.~E {\bf 91},  022817  (2015).

\bibitem{tinsley:2012} M. R. Tinsley and K. Showalter, 
Nature Physics \textbf {8}, 662 (2012).

\bibitem{hagerstrom:2012} A. M. Hagerstrom, E. Thomas, R. Roy, P. H\"ovel, I. Omelchenko and E. Sch\"oll, 
Nature Physics \textbf {8}, 658 (2012).

\bibitem{martens:2013}
E.~A. Martens, S. Thutupalli, A. Fourri{\`e}re, and O. Hallatschek, Proc. Nat.
  Acad. Sciences {\bf 110},  10563  (2013).


\bibitem{LAR13}
L. Larger, B. Penkovsky, and Y. L. Maistrenko,
Phys. Rev. Lett. \textbf {111}, 054103 (2013).
  
\bibitem{SCH14a}
L. Schmidt, K. Sch{\"o}nleber, K. Krischer, and V. Garcia-Morales,  
Chaos \textbf {24} 013102 (2014).

\bibitem{WIC13}
M. Wickramasinghe and I.~Z. Kiss, PLoS ONE {\bf 8},  e80586  (2013).

\bibitem{panaggio:2014} M. J. Panaggio and D. M. Abrams, 
Nonlinearity \textbf{28}, R67 (2015).

\bibitem{ma:2010} R. Ma, J. Wang and Z. Liu, 
Europhys. Lett. \textbf{91}, 40006 (2010).

\bibitem{davidenko:1992} J. M. Davidenko, A. V. Pertsov, R. Salomonsz, W. Baxter, and J. Jalife, 
Nature \textbf{355} (6358), 349 (1992).


\bibitem{murray:1993} J. D. Murray, 
\textit{Mathematical Biology}, Springer-Verlang, Berlin 1993.

\bibitem{may:2001} R. M. May, 
\textit{Stability and Complexity in Model Ecosystems}, 
Princeton University Press, Princeton, 2001.

\bibitem{llv:1999} A. Provata, G. Nicolis, 
and F. Baras, J.~Chem.~Phys. \textbf{110}, 8361 (1999).

\bibitem{frachebourg:1996} L. Frachebourg, P. L. Krapivsky, and E. Ben-Naim, 
Phys. Rev.~E \textbf{54}, 6186 (1996).

\bibitem{tsekouras:2001} G.A. Tsekouras and A. Provata, 
Phys. Rev.~E \textbf{65}, 056602 (2001).

\bibitem{imbihl:1995} N. Khrustova, G. Veser, A. Mikhailov, and R. Imbihl,
Phys. Rev. Lett. \textbf{75}, 3564 (1995).

\bibitem{imbihl2:1995} R. Imbihl and G. Ertl 
Chem. Rev. \textbf{95}, 697 (1995).
  
\bibitem{noussiou:2007} V. K. Noussiou, and A. Provata,
Surface Science \textbf{601}, 2941 (2007).

\bibitem{anderson:1995} A. B. Anderson and E. Grantscharova, 
J.~Phys. Chem. \textbf{99}, 9149 (1995).

\bibitem{govender:2008} N.S. Govender, F. G. Botes, M. H. J. M. de Croon, and J.C. Schouten,
J.~Catalysis \textbf{260}, 254 (2008).

\bibitem{ertl:1994} G. Ertl,
 Surface Science \textbf{299/300}, 742 (1994).


\bibitem{epstein:1998}  I. R. Epstein and J. A. and Pojman, 
\textit{An introduction to nonlinear chemical dynamics: oscillations, waves, patterns, and chaos}, 
Oxford University Press, New York, 1998.

\bibitem{nicolis:1977} G. Nicolis and I. Prigogine, 
\textit{Self-Organization in Nonequilibrium Systems}, Wiley, New York, 1977.

\bibitem{anderson:1991} R. M. Anderson and R. M. May,  \textit{Infectious Diseases of Humans}, 
 Oxford University Press, Oxford, (1991).

\bibitem{delitala:2004} M. Delitala, 
Math. and Comp. Modelling \textbf{39}, 1 (2004).
     
\bibitem{pastor:2001} R. Pastor-Satorras and A. Vespignani,
Phys. Rev. Lett. \textbf{86}, 3200 (2001).

\bibitem{llc:2002} A. V. Shabunin, F. Baras. and A. Provata, 
Phys. Rev.~E \textbf{ 66},   036219 (2002).

\bibitem{gonzalez:2014} J. C. Gonzalez-Avella, M. G. Cosenza, and M. San Miguel,
Physica A \textbf{399}, 24 (2014).

\bibitem{axelrod:1997} R. Axelrod, 
J.~Conflict Res. \textbf{41}, 203 (1997).

\bibitem{deffuant:2000} G. Deffuant, D. Neau, F. Amblard, and G. Weisbuch,
Adv. Comp. Sys. \textbf{3}, 87 (2000).

\bibitem{batty:2008} M. Batty, 
Science \textbf{319}, 769 (2008).
%
 \bibitem{sole:1999} R. V. Sol\'e, S.C. Manrubia, M. Benton, S. Kauffman, and B. Per, 
 Trends in Ecology and Evolution \textbf{14}, 156 (1999).
%
 \bibitem{campos:2013} P. R. A. Campos, A. Rosas, V. M. de Oliveira, and  M. A. F. Gomes  
PLoS ONE \textbf{8}, e66495 (2013). 
%
\bibitem{buchmann:2013} C. M. Buchmann, F. M. Schurr, R. Nathan, and F. Jeltsch, 
Ecological Informatics \textbf{14}, 90 (2013).
%
\bibitem{wallace:1994} R. Wallace,
Environment \& Planning A \textbf{26}, 767 (1994).
%

\bibitem{barthelemy:2005} M. Barthelemy, A. Barrat, R. Pastor-Satorras,  and A. Vespignani,
J.~Theor. Biol. \textbf{235}, 275 (2005).

\bibitem{zhdanov:2002} V. P. Zhdanov,
Surface Science Reports \textbf{45}, 233 (2002). 

\bibitem{panagakou:2013} E. Panagakou, G. C. Boulougouris, and A. Provata,
Eur. Phys. J.~B \textbf {86}, 277 (2013).

\bibitem{provata:2014} A. Provata, and E. Panagakou
\textit{submitted}  (2014).

\bibitem{OME10a}
O.~E. Omel'chenko, M. Wolfrum, and Y.~L. Maistrenko, Phys. Rev.~E {\bf 81},
  065201(R)  (2010).

\bibitem{SET08} G. C. Sethia, A. Sen, and F. M. Atay  
Phys. Rev. Lett. \textbf{100}, 144102
   (2008).

\bibitem{MAI14}
Y.~L. Maistrenko, A. Vasylenko, O. Sudakov, R. Levchenko, and V.~L. Maistrenko (2014)
 arXiv:1402.1363v1.

\bibitem{OME15}
I. Omelchenko, A. Provata, J. Hizanidis, E. Sch{\"o}ll, and P. H{\"o}vel, Phys.
  Rev. E {\bf 91},  022917  (2015).

\bibitem{feder:1988} J. Feder, \textit{Fractals}, Plenum Press, New York, 1988.

\end{thebibliography}
\end{document}